\documentclass[11pt]{article}
\usepackage[a4paper,margin=2.7cm]{geometry}
\usepackage{amsmath,amssymb}
\usepackage{graphicx}
\usepackage{booktabs}
\usepackage[round]{natbib}
\usepackage[colorlinks=true,linkcolor=blue!60!black,citecolor=blue!60!black,urlcolor=blue!60!black]{hyperref}
\usepackage{xcolor}
\usepackage{longtable}
\usepackage{float}
\usepackage{array}

\graphicspath{{figs/}}
\newcommand{\astar}{\alpha^{*}}

\title{Message capacity and claim wording set the transition points of collective truth-finding in language-model networks}
\author{Makoto Fukushima\\[3pt]
{\normalsize Research Division, Honda Research Institute Japan Co., Ltd.,
Wako, Saitama, Japan}\\[2pt]
{\small Correspondence: \texttt{makoto.fukushima@jp.honda-ri.com}}}
\date{}

\begin{document}
\maketitle

\begin{abstract}
Whether human or large language model (LLM), an agent in a discussion reads
only a few of the others' contributions, bounded by cognition, context, or
cost. LLM collectives can settle on a wrong consensus even when a majority
starts out correct; we ask how far that reading bound alone decides the
outcome. We model the bound with one number, the message capacity, which
sets how many of the others' messages an agent reads, and generate the
communication network from it. Over 31{,}824 randomized queries, we found
that an 8-billion-parameter model's judgment of a claim effectively reduces
to a logistic function of a weighted sum of its inbox, the update rule of a
stochastic binary neuron with divisively normalized weights. From these
weights and the network's degree statistics alone, the wrong consensus
should become unreachable from any start once agents read, on average,
fewer than 6.4 of their 31 sources. In 1{,}414 episodes with assigned
starts the prediction failed: the correct side won in fewer than 50\% of
episodes from every start, and in only 28--45\% when 75\% of agents started
correct. The failure traces to the field, the threshold that a claim's wording sets for the agent's
answer before any message is read: the experimental claims' fields lay below the calibration
mean, and with each claim's own field the same weights reproduce the
outcomes. Reversing the wording showed that the threshold follows what a claim asserts, not whether it is true. On a second 8B model the pipeline predicts
claim-dependent bistability; transition points appeared where computed, and
an eight-claim calibration matched in 15 of 16 conditions. At 70B the
assertion bias is not detected. Thus a collective's fate is largely set by
two single-agent measurements: the threshold a claim's wording sets, and the message capacity that sets
the transition point.
\end{abstract}

\section*{Significance statement}

Language-model agents work in groups that debate and settle on answers, and
such groups can end up wrong. A group's fate is largely set by two quantities
measurable on a single agent. An agent's judgment follows the input-output
rule of a stochastic binary neuron, each message counting less as the inbox
grows, so the group's transition point follows from one equation given how
many sources each member reads. The second quantity is the threshold a
claim's wording sets for an agent's answer before it reads any message.
Across three models, collectives followed the assertion rather than the
truth in one and crossed predicted transition points in another. AI
collectives can be assessed, and designed, before they run.

\section{Introduction}\label{sec:intro}

Whether human or large language model (LLM), an agent in a discussion
processes only a few of the others' contributions. For people the bound is
cognitive. Working memory holds only about four items at once
\citep{Cowan2001,LuckVogel1997}, and selective attention narrows the intake
further: a listener in a crowded room follows one conversation and retains
almost nothing of the others \citep{Cherry1953}, because unattended streams
are attenuated before they reach the capacity-limited stage
\citep{Treisman1964}. In online discussions the same limit appears as
divided attention capping how many contributions a participant actually
processes \citep{HodasLerman2012}. For LLM agents the bound is set by
context windows, per-token costs, and the sampling policies of agent
frameworks.
Groups of LLM agents that exchange messages and update binary opinions are
now studied as statistical-mechanical systems \citep{El2026}, and whether
such a group finds the truth is not guaranteed: members who each read only
part of the discussion may carry the group to a consensus that most of them
did not start with. The question this paper
asks is to what extent that reading bound alone, together with the response
of a single agent measured in isolation, decides where a collective ends up.

Two ingredients make the question tractable. The first turns the bound into
a network. In a model of synaptic resource allocation \citep{SciRepCrowding},
a receiving unit that has already accepted $r$ inputs accepts a further one
with probability $e^{-\alpha r}$; the number of accepted inputs then grows
only logarithmically in the candidate pool, the mean in-degree is
$(\log N)/\alpha$ with bounded variance, the full in-degree distribution
$P_{\alpha,N}(k)$ obeys a closed recursion, and the distribution is
invariant under reordering of the candidates. One parameter, $\alpha$ (the
crowding parameter of that model), compresses everything about the acceptance process that matters for degree
statistics: applied to a discussion, it is the message capacity of an agent,
and larger $\alpha$ means fewer messages read. Throughout, $\alpha$
describes the crowding of a finite intake budget at the level of whole
messages; it is unrelated to the attention mechanism inside transformer
architectures, and we avoid the word ``attention'' for this reason.

The second ingredient describes the agent. A model that reads $k$ messages
and then judges a claim could respond in countless ways; a theory needs a
rule simple enough to iterate, and the rule must be measured rather than
assumed. We take the agent to be a stochastic binary neuron: its probability
of siding with the correct answer is a logistic function of a weighted sum
of its inbox, one weight per correct-side message and one per wrong-side
message, with weights that may depend on inbox size. We do not assume this
form; we measure it on the experimental agent, and check it against the
largest public dataset of LLM communities, before any collective is run.
With a measured neuron and a
crowding-generated network, the collective becomes a network of binary
neurons whose degree statistics are known in closed form, and mean-field
theory yields a one-dimensional map whose fixed points are the possible
consensus states.

The two ingredients together pose a quantitative question. Consider a
population in which a fraction $x_0$ of agents initially holds the correct
answer to a binary question. After several rounds of exchange the population
typically reaches a correct or a wrong consensus; the probability of the
former after a fixed number of rounds, $q(x_0)$, is the finite-horizon
analogue of the committor of the collective dynamics, the probability of
reaching one end state rather than the other from a given start. Mean-field
theory on the generated networks predicts that the basin of the wrong
consensus disappears entirely above a critical crowding level $\astar$: for
$\alpha > \astar$, even small correct minorities propagate. Our question is
to what extent $\astar$ can be computed in advance, from the degree
statistics that $\alpha$ induces and from response coefficients measured on
single agents in isolation, with no quantity fitted to collective runs. The
nearest precedents come from LLM naming games, where emergent conventions
and collective bias are established phenomena \citep{Ashery2025, Flint2026}:
a mean-field consensus condition has been assembled from single-token
acceptance rates measured on isolated models \citep{DeNobili2026}, and a
mean-field account of population-size-dependent collective bias from
logit-extracted single-agent policies \citep{Flint2026}. These constructions
share our logic, but they concern coordination on a convention (no correct
side, no committor) under fully mixed or imposed interaction structures. In
the truth-finding setting, \citet{El2026} showed that a kinetic Ising model
fitted to one-step transitions describes more than 10{,}000 communities
($N = 32$, eight rounds) and generalizes to unseen graphs; there, as in most
multi-agent opinion studies, the graph is supplied by the experimenter and
the rule is fitted to the collective data. No previous study has, to our
knowledge, generated the interaction network of an LLM collective from a
crowding process, or predicted the basin boundary of a truth-finding
collective, the critical initial fraction of correct agents, from
single-agent measurements completed before the collective ran.

Two families of confound make this comparison easy to get wrong, and both
shaped our design. First, in observational data the initial fraction $x_0$
is not an experimental dial: where question difficulty drives both the
initial fraction and the outcome, the apparent committor is a difficulty
curve and the transition point (the tipping point of the social-dynamics
literature) is unidentified. We examine this concern on the largest public
dataset, and our collective experiment assigns initial opinions
exogenously. Second, changing the network changes the
composition of each agent's prompt, and LLMs respond to prompt composition
regardless of any network; effects of this kind reproduce in a single agent
with a single query. We therefore measure the single-agent response function
first, predict the collective outcome from it, and treat any residual,
quantified by replaying every realized prompt through the measured response
function, as the collective contribution proper.

The bound itself is not particular to machines, and the theory's inputs (a response threshold, per-message weights and their attenuation, an intake
bound)
are behavioral quantities with human counterparts. A theory that turns an
intake bound into a collective outcome is therefore, at least potentially, a
statement about committees and juries as well as about machine collectives.
In human groups, however, neither the bound nor the network it induces can
be measured or controlled; in LLM collectives, both can, which is why the
test is made here.

The paper proceeds in three steps: a theory of $\astar$ from a reduced map
(Section~\ref{sec:theory}), the measurements that give the theory its
coefficients, from a reanalysis of the public dataset and a direct measurement
of the experimental agent (Section~\ref{sec:measurement}),
and a collective experiment run against a prediction registered before any
collective episode (Section~\ref{sec:collective}), followed by the
Discussion. Whether the transition point of an LLM collective can be
computed before it runs, and what else has to be measured when it cannot,
is the question the rest of the paper addresses. A glossary of the terms
used below is given in SI Section~S1.

\begin{figure}[t]
\centering
\includegraphics[width=\textwidth]{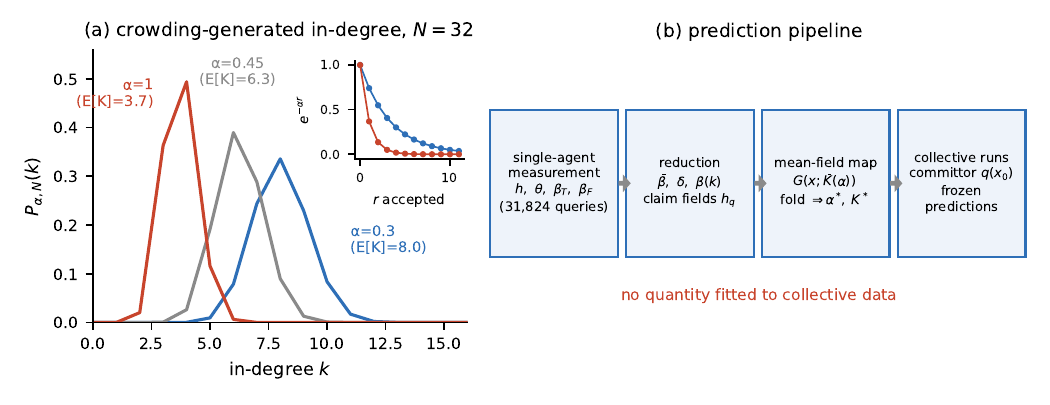}
\caption{\textbf{Crowding-generated networks and the prediction pipeline.}
(a)~In-degree distribution $P_{\alpha,N}(k)$ generated by the acceptance
process $e^{-\alpha r}$ (inset) at three crowding levels, $N=32$.
(b)~Every quantity used to predict the collective outcome is measured on
single agents or computed from the reduced map; nothing is fitted to
collective data.}
\label{fig:framework}
\end{figure}

\section{Theory: a critical crowding level from a reduced map}\label{sec:theory}

The theory assembles two ingredients: the crowding-generated network
(Section~\ref{sec:netgen}) and the measured single-agent rule
(Section~\ref{sec:reduction}). It then asks at what capacity the
wrong-consensus basin disappears (Section~\ref{sec:critical}).

\subsection{Network generation}\label{sec:netgen}

We generate the network by the process of \citet{SciRepCrowding}. Each
receiving agent processes the other $N-1$ agents in arbitrary order. Having
already accepted $r$ sources, it accepts the next with probability
$e^{-\alpha r}$; the first candidate is always accepted. The in-degree
distribution $P_{\alpha,N}(k)$ then follows the recursion
$P_{m+1}(r) = P_m(r)\,(1-e^{-\alpha r}) + P_m(r-1)\,e^{-\alpha (r-1)}$,
with mean $E[K] = (\log N)/\alpha + O(1)$. New here is only the closure.
The deterministic continuum limit gives the finite-$N$ form
\begin{equation}
\bar K(\alpha) = \frac{1}{\alpha}\,\ln\!\bigl(1 + \alpha (N-1)\bigr),
\end{equation}
which underestimates the exact mean by 3\% ($\alpha = 0.3$) to 11\%
($\alpha = 2$) at $N = 32$. Where precision matters we therefore invert the
exact recursion instead. Note that Eq.~(1) is a large-$N$,
moderate-$\alpha$ form: as $\alpha \to \infty$ it tends to zero, whereas
the exact mean tends to one because the first candidate is always accepted.
One property of the process matters for the experimental design. Degree
statistics are invariant under reordering of the candidates, so
manipulations of \emph{who} is read change nothing in $P(k)$. Thus,
order-based conditions can be compared at exactly matched communication
budgets.

\subsection{Single-agent response and its reduction}\label{sec:reduction}

We take the agent's rule from the measurements of
Section~\ref{sec:measurement}. An agent receiving $l$ correct-side and
$k-l$ wrong-side messages adopts the correct side with probability
\[
\sigma\bigl(c_0 + \theta\,\mathbf{1}[\text{own state correct}] + \beta_T l +
\beta_F (k-l)\bigr),
\]
where $\sigma$ is the logistic function and the
per-message weights are attenuated by inbox size as $\beta g(k)$
(Section~\ref{sec:measurement}). The measured unit is therefore equivalent
to a stochastic binary neuron.

Three combinations of its coefficients organize everything that follows.
The first is the symmetry-breaking field $h = c_0 + \theta/2$, measured
per claim (the persistence indicator, which takes the values 0 and 1,
contributes a constant part $\theta/2$ to the field and a symmetric part
that does not). In the language of the binary neuron, $h$ is the unit's
threshold with its sign reversed. The agent answers on the correct side
when its weighted input exceeds $-h$, so $h$ is the log-odds of the correct
side with an empty inbox. A positive $h$ means that the claim's wording
alone lowers the threshold for the correct side; a negative $h$ means that
it raises it. The threshold shifts the logistic response along its input
axis, whereas the steepness of the response is set by the message weights,
the gains. We refer to $h$ as the threshold that the claim's wording sets,
and we report it as $h$ throughout. The second combination is the
per-input swing $\bar\beta = (\beta_T - \beta_F)/2$, the average push of
one message toward its own side. The third is the per-input drift
$\delta = (\beta_T + \beta_F)/2$, the push that any message adds regardless
of its side.

How does a population of such units evolve? In words, the population
dynamics is a map that takes the fraction $x$ of correct-side agents in one
round and returns the expected fraction in the next: an agent with $k$
sources sees each of them on the correct side with probability $x$ and
responds through the measured rule. A fixed point of the map ($F(x) = x$)
is a state the collective can settle into. A stable fixed point is a
consensus, and an unstable one lying between two stable ones is the
transition point that separates their basins. Under the heterogeneous
mean-field assumption (each of an agent's $k$ inputs is correct
independently with probability $x$) the map is
\begin{equation}
F(x) = \sum_{k} P_{\alpha,N}(k) \sum_{l=0}^{k} \binom{k}{l}
x^{l} (1-x)^{k-l} \bigl[\, x\,\sigma(u_{1}) + (1-x)\,\sigma(u_{0})
\bigr],
\label{eq:hmfmap}
\end{equation}
with $u_{s} = c_0 + \theta s + \beta_T l + \beta_F (k-l)$. The interior
unstable fixed point of $F$ separates the two consensus basins.

This construction is the mean-field analysis of the generating model
\citep{SciRepCrowding} with one substitution: the hard-threshold unit
assumed there is replaced by the measured logistic unit, of which the
hard-threshold unit is the infinite-gain limit. The outcome observable
$q(x_0)$ is likewise inherited from that analysis, which computed the
committor proper by an absorbing Markov-chain closure; here it is read
after eight rounds (Methods). Note that no hard-threshold nonlinearity is
assumed anywhere below. Bistability is a property of the fixed-point
structure of the S-shaped map, an unstable interior fixed point separating
two stable ones. The substitution also removes the structural
discontinuity that hard-threshold updates produce at truth-weight 1 (equal
per-message weight for correct-side and wrong-side sources;
Section~\ref{sec:measurement}).

Exact numerics iterate Eq.~(\ref{eq:hmfmap}) directly. To obtain a form
that can be read by eye, we close the degree at $\bar K(\alpha)$ and smooth
the binomial with the standard Gaussian--logistic approximation, which
yields the reduced map
\begin{equation}
G(x;K) = x\,\sigma(z_+) + (1-x)\,\sigma(z_0), \qquad
z_s = \frac{(h \pm \theta/2) + K\delta + 2\bar\beta K (x - \tfrac12)}
{\sqrt{1 + \tfrac{\pi}{8}\, 4\bar\beta^2 K x(1-x)}}.
\end{equation}
A degree-dependent coupling $g(k)$, for example the divisive normalization
$\beta/(1+\gamma k)$, one of the two attenuation forms our data support,
enters by multiplying $\bar\beta$ and $\delta$ and leaves the form
unchanged.

\subsection{The critical capacity}\label{sec:critical}

As $\alpha$ grows and each agent reads fewer messages, the interior
unstable fixed point moves toward the wrong-consensus fixed point. At
$\astar$ the two merge and disappear, and beyond it only the correct
consensus remains. Formally, $\astar$ is the capacity at which the interior
unstable fixed point of $G$ undergoes a fold ($G = x$ and $G' = 1$
simultaneously). We keep the term critical crowding level for $\astar$ for
continuity with the generating model. The transition is, however, a
saddle-node bifurcation of the mean-field map, corresponding to the
disappearance of a metastable basin (a spinodal), and the Discussion
returns to the distinction. Operationally, the fold defines a critical
degree $K^*$, and $\astar$ solves $\bar K(\alpha) = K^*$.

The mechanism is visible in the linear gain at the symmetric point,
\begin{equation}
\lambda(K) = \bigl[\sigma(z_+) - \sigma(z_0)\bigr]
+ \frac{2\bar\beta K\,\bar\sigma'}{\sqrt{1 + \tfrac{\pi}{8}\bar\beta^2 K}},
\end{equation}
where $\bar\sigma' = [\sigma'(z_+) + \sigma'(z_0)]/2$ and every term is
evaluated at $x = 1/2$. The first term is a persistence gain, carried by
$\theta$. The second is a social-amplification gain that, for constant
coupling, grows only as $\sqrt{K}$, because input fluctuations smooth the
response. In the symmetric case ($h = 0$ and $\delta = 0$) bistability
requires $\lambda > 1$. With a nonzero field, however, the gain at the
symmetric point locates the mechanism but is neither necessary nor
sufficient for bistability, and it misplaces $\astar$ by 30--90\% at
realistic fields. All quantitative predictions below therefore come from
the fold of $G$.

One further structural fact falls out of the reduction. With the measured
$\delta < 0$, the effective field $h + K\delta\,g(K)$ is screened as degree
grows; with constant coupling it crosses zero near $K \approx 10$ for the
coefficients measured in Section~\ref{sec:measurement} under the
collective experiment's prompt (variant A). This is why bistability
survives at low $\alpha$ despite $h > 0$. Under divisive coupling the
screening saturates, $K\delta\,g(K) \to \delta/\gamma$, and over the mean
degrees realized here ($K \approx 3$--$10$) the factor $K g(K)$ varies only
between 0.82 and 1.0. Thus, the effective field, and with it the position
of the transition point, depends only weakly on $\alpha$. The weak
$\alpha$-dependence of the qwen3:8b transition points
(Section~\ref{sec:generality}) is consistent with this; the zero crossing
itself is not tested separately.

\begin{figure}[t]
\centering
\includegraphics[width=\textwidth]{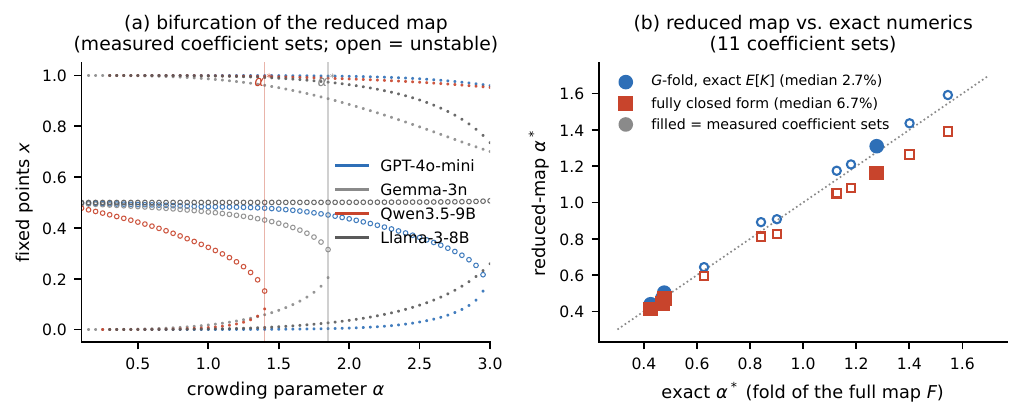}
\caption{\textbf{Critical crowding level from the reduced map.} (a)~Fixed points of the
reduced map $G$ as functions of $\alpha$ for the four measured coefficient
sets of \citet{El2026}; the interior unstable branch (open symbols) vanishes
at $\astar$. (b)~Validation of the two closure levels against the fold of
the full map $F$ over 11 coefficient sets (filled symbols: coefficient sets
measured in this study).}
\label{fig:theory}
\end{figure}

How accurate is the closure? Against exact numerics (the full $P(k)$ sum,
fold located by bisection in $\alpha$), it is accurate to a median 2.7\%
over a 3-decade parameter grid and both measured coefficient sets when the
exact $E[K]$ is inverted, and to 6.7\% in the fully closed form. The worst
cases are 6.0\% (exact $E[K]$; the synthetic set $h = 0.7$, $\theta = 0.6$,
$\bar\beta = 1$) and 9.9\% (closed form; $h = 0.3$, $\theta = 2$,
$\bar\beta = 1$) (Fig.~\ref{fig:theory}; SI). For the measured variant-A
coefficients the prediction is $\astar = 0.435$, $K^* = 6.4$, as registered
(the fold conditions solved to full precision give 0.439 and $K^* = 6.36$,
inside the registered tolerance). In words, the wrong-consensus basin is
predicted to disappear when agents accept, on average, fewer than 6.4 of
their 31 possible sources. Thus, the critical crowding level of the
collective is, to this accuracy, computable from single-agent coefficients
through the reduced map.

\section{Measuring the response function of the experimental agent}
\label{sec:measurement}

\subsection{What the public dataset provides, and what it cannot test}

The natural first test requires no new data. \citet{El2026} released full
logs of 6{,}400 objective-question communities (four models, 40 questions,
ten graphs per question, four episodes) \citep{El2026data}. We therefore
asked whether their apparent committor $q(x_0)$ identifies a transition
point. It does not, for a reason that constrains any design in which the
initial fraction is not assigned. Across the four models, between 50\% and
99.5\% of the variance of $x_0$ lies between questions, and within
questions the $x_0$--outcome correlation drops to 0.04--0.27. Question
difficulty thus drives both the initial fraction and the outcome, so the
pooled $q(x_0)$ curve is a difficulty curve, and its 0.5-crossing is not a
property of the collective dynamics. Accordingly, a question-clustered
bootstrap widens the intervals on the fitted crossing 4--10 fold, after
which no difference among models or graph families is resolved. Two
further obstacles are structural. The released graphs have nearly
degenerate degree distributions, so the degree-statistics content of any
mean-field prediction is untestable there; and under the hard-threshold
update rule assumed in earlier work the predicted crossing jumps
discontinuously at truth-weight 1, which reduces the comparison to a single
bit (SI Section~S2).

The logs do yield the microscopic coefficients. We regressed each agent's
next-round correctness on the composition of its inbox (43{,}500
observations per model, question-clustered errors). The truth asymmetry
$w_T$, the relative per-message weight of correct-side sources previously
estimated at 1.16--1.71 from Ising couplings, is indistinguishable from 1
in all four models (GPT-4o-mini: 1.055, 95\% CI $[0.996, 1.114]$). A
persistence term $\theta$ and an individual field $c_0$, in contrast, are
robustly nonzero. The asymmetry that breaks the symmetric degeneracy of the
mean-field map therefore resides in the field $h = c_0 + \theta/2$, not in
message weighting, which is what Section~\ref{sec:theory} assumes. Thus,
public data fix the form of the theory, but they cannot test its
collective prediction.

\subsection{The experimental agent}

A theory needs the agent's rule measured rather than assumed, and this
section is that measurement. The collective experiment uses llama3.1:8b
(4-bit Q4\_K\_M quantization, temperature 0.7, snap-judgement prompt
matching El et al.'s protocol). We measured its response function directly
with 31{,}824 single-shot queries, in which the inbox composition ($k$
messages, $l$ of them correct-side), the A/B label assignment, the message
order, and an optional explicit own-state line were randomized by design.
Claims came from CLIMATE-FEVER \citep{Diggelmann2020}. A two-stage
calibration under the bare judgment prompt selected 17 claims with
per-claim field $|h_q| < 0.4$. Under the full experimental scaffold
(persona and inbox lines), however, the fields did not transfer: the
per-claim intercepts re-expanded to a range of 3.5 logits with mean 0.94.
This is why the claims of the collective experiment were calibrated again
under the full scaffold (Methods), and, as Section~\ref{sec:results} shows,
why the field rather than the message weights turns out to be the quantity
that fails to carry across claim sets. Note that 16 of the 17 selected
claims are REFUTES (claims labeled false in CLIMATE-FEVER). All results are
therefore stated in correct-side/wrong-side terms, and the
truth-versus-polarity distinction is explicitly not identified by this
claim set: a diagnostic regression shows that TRUE-side messages are more
persuasive on average (ratio 1.44 $[1.25, 1.65]$), which accounts for the
apparent $w_T = 0.74$--$0.89 < 1$.

\begin{figure}[t]
\centering
\includegraphics[width=\textwidth]{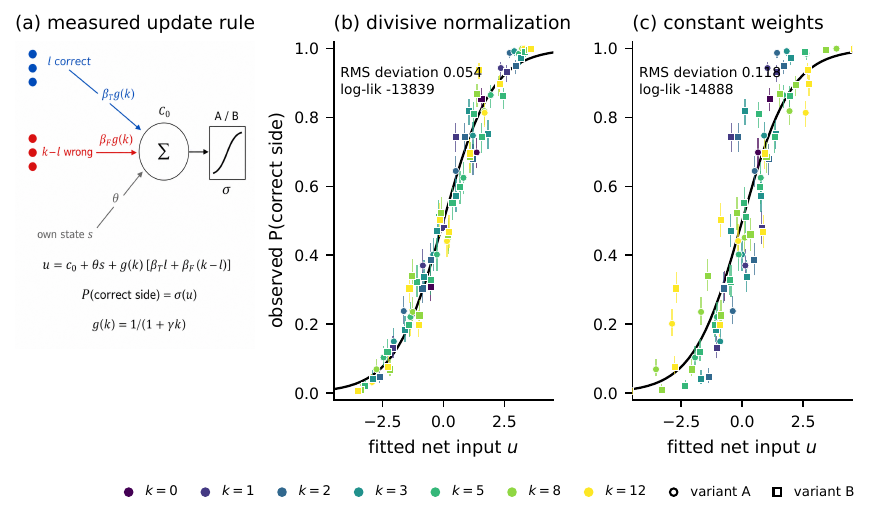}
\caption{\textbf{Observed correct-side rate against fitted net input under divisive and constant message weights.}
(a)~The measured update rule: $l$ correct-side and $k-l$ wrong-side messages
enter with weights $\beta_T g(k)$ and $\beta_F g(k)$, the own-state indicator
$s \in \{0,1\}$ (variant B only) with weight $\theta$, and the field $c_0$;
the choice probability is $\sigma(u)$ with $g(k) = 1/(1+\gamma k)$,
$\gamma = 0.9$. (b)~Observed correct-side rate in each of the 78 design cells
of the 31{,}824 randomized queries (variant $\times$ $k$ $\times$ $l$
$\times$ $s$; 408 queries per cell; Wilson 95\% intervals) against $u$, the
logit of the model's mean predicted probability in the cell under divisive
normalization with claim fixed effects. Color gives inbox size $k$; circles
variant A, squares variant B. (c)~The same cells under constant
weights ($g \equiv 1$): large-$k$ cells lie between the curve and 0.5 and
small-$k$ cells beyond it. RMS deviation of cell rates from $\sigma(u)$: 0.054 in
(b), 0.118 in (c).}
\label{fig:neuron}
\end{figure}

Three measured facts carry the predictions (Fig.~\ref{fig:neuron}; per-$k$
coefficients and interval estimates in SI Fig.~S1).

The first is the set of coefficients. With no line in the prompt stating
the agent's own current answer (variant A, the protocol of the collective
experiment), the field is $h = 0.725$ $[0.516, 0.915]$. With such a line
(variant B), the persistence term is $\theta = 1.909$ $[1.723, 2.106]$.
Adding one line thus raises $\theta$ from absent to about 1.9. This is
consistent with the persistence measured in observational data being a
persona-level carryover rather than an in-prompt self-input. Fitted jointly
with claim fixed effects anchored by the empty-inbox cells, the per-message
coefficients fall from $\beta_T = 2.11$ $[1.58, 2.96]$ and
$\beta_F = -1.53$ $[-2.03, -1.12]$ at $k = 1$ to $0.23$ $[0.18, 0.27]$ and
$-0.39$ $[-0.48, -0.33]$ at $k = 12$ (SI Fig.~S1). Note that separate fits
at fixed $k$ do not identify these coefficients (Methods).

The second is the load attenuation: each additional message weakens the
influence of every message. To identify its form, we compared candidate
forms by cross-validation split by claim. Forms that depend on inbox size
were favored, load decay $e^{-0.14k}$ and divisive normalization
$\beta/(1+0.9k)$ being indistinguishable, over the position-dependent form
$e^{-\alpha_{\mathrm{pos}} r}$ closest to the theory's acceptance kernel.
The theory's kernel is thus implemented by the exogenous router, not by the
model's internal weighting, under this protocol, in which message order is
randomized. A deployment that orders messages by recency or salience could
reinstate an internal position dependence. This demotion was fixed in the
design before any collective run, and the internal decay enters the theory
only through $\beta(k)$. We quantified the reduction itself in two ways: by
claim-clustered cross-validated log-likelihood ($-0.483$ nats per
observation for the two retained forms, against $-0.509$ for a
constant-weight model with the same claim fields and self term and $-0.693$
for chance), and by the calibration of the fitted $\sigma$ on every
replayed collective transition (expected calibration error 0.03;
Section~\ref{sec:collective}). The same reduction is visible cell by cell
(Fig.~\ref{fig:neuron}b,c): the RMS deviation of the 78 cell rates from the
logistic curve is 0.054 with divisive weights against 0.118 with constant
weights. This comparison is in-sample; the cross-validated log-likelihoods
above are the out-of-sample statement.

The third is the prediction. Pushing the measured coefficients through
Section~\ref{sec:theory} gives, as folds of the full map $F$,
$\astar = 0.43$--$0.48$ for variant A \emph{under every reduction of the
$k$-dependence} (constant 0.43, or 0.435 by the reduced map as registered;
divisive 0.48; jointly identified per-$k$ coefficients 0.46). For variant B
the reductions disagree: $\astar = 0.47$ (constant) versus 1.28 (divisive).
This disagreement is an experimental resource, because an arm run under
variant B between $\alpha = 0.6$ and 1.3 discriminates the coupling rule
regardless of outcome. Thus, the response function of a single agent,
measured in isolation, fixes every number the collective experiment is
asked to reproduce.

\section{Collective experiment}\label{sec:collective}

\subsection{Design (pre-registered before any collective run)}

With the network from $\alpha$ and the neuron from
Section~\ref{sec:measurement}, the collective experiment tests the
prediction that follows from them. Each episode is one collective run. A
network is generated by the acceptance process at capacity $\alpha$,
initial opinions are assigned exogenously ($\lceil 32 x_0 \rceil$ agents
seeded correct-side), and the dynamics follow El et al.'s protocol for
eight rounds (one broadcast message per agent per round conditioned on its
current stance, Markovian inboxes, snap-judgement opinion resampling).

Two arms were run. The main arm (variant A prompts) covers
$\alpha \in \{0.20, 0.30, 0.38,\allowbreak 0.45, 0.55, 1.00\}$ and
$x_0 \in \{4,\dots,24\}/32$, adaptively concentrated where $q$ crosses
$\tfrac12$ and, above the predicted $\astar$, at low $x_0$ where the
collapse of the wrong basin is most visible. The discrimination arm
(variant B prompts) runs $\alpha \in \{0.60, 0.90, 1.30\}$, the range over
which the two coupling rules of Section~\ref{sec:measurement} disagree.

Three analyses accompany the runs. A rewiring control
(in-degree-preserving shuffle with rewiring fraction $\rho = 1$, i.e., all
sources resampled, at $\alpha = 0.30$) checks that outcomes are
reproducible when every receiver's sources are redrawn at fixed in-degree;
this is a test of reproducibility under the mean-field premise rather than
of its sufficiency (Section~\ref{sec:results}). A yoked one-step analysis
replays every realized inbox through the Section~\ref{sec:measurement}
response function, so that any collective deviation from iterated
single-agent behavior is measured rather than assumed absent. A surrogate
rollout simulates the measured per-$k$ rule for eight rounds on the same
graphs and seeds and provides the finite-size prediction band.

Decision criteria were fixed in advance. The empirical
$\astar_{\mathrm{emp}}$ is the capacity at which the wrong-consensus rate
from $x_0 \le 8/32$, after eight rounds, falls through 20\%, and the
discrimination arm's verdict is read from which $\alpha$ levels retain a
wrong basin. Outcomes are read after eight rounds: an episode ends in a
correct-side consensus when $x_8 \ge 0.6$ and in a wrong-side consensus
when $x_8 \le 0.4$. The criterion is operational. Among decided episodes
the final fraction lies beyond $|x_8 - 0.5| > 0.4$ in 63\% of the main
arm's episodes (median wrong-side end state $x_8 = 0.03$) and in 97--99\%
of the qwen3:8b campaigns of Section~\ref{sec:generality}, so the labels
describe strongly polarized end states. The sensitivity of every result to
the criterion ($\delta \in \{0.05, 0.2\}$) is reported in the SI.

\subsection{The pre-registered prediction fails}\label{sec:results}

We first tested the pre-registered prediction directly. Across all 1{,}414
episodes (zero parse failures), the probability of a correct consensus
never reached 0.5 in any $(\alpha, x_0)$ cell (Fig.~\ref{fig:collective}a).
The collectives converged on the wrong consensus from most initial
conditions. Even when 24 of the 32 agents started on the correct side, the
correct side won in only 28--45\% of episodes (wrong side 33--73\%,
undecided 0--28\%). The wrong-consensus rate from low $x_0$ did fall as
$\alpha$ increased, from 0.98 at $\alpha = 0.20$ to 0.75 at
$\alpha = 1.00$, which is the direction the theory predicts. However, it
never crossed the pre-registered 20\% criterion, so no empirical critical
crowding level exists in the tested window. Thus, the prediction
$\astar = 0.435 \pm 0.03$ is rejected (claim-clustered bootstrap, 0 of
2{,}000 resamples crossing).

\subsection{The failure traces to the threshold, not to the message weights}

Why did the prediction fail? The theory takes two kinds of single-agent
input: the message weights, which were pooled over the 17 calibration
claims, and the field, which the blind rule also inherited from the
calibration set. To find out which of the two failed, we replayed every
realized inbox through the response function of
Section~\ref{sec:measurement}, keeping the pooled weights but fitting the
field to each experimental claim. This replay reproduced the observed
transitions without systematic error: the Brier score was 0.10--0.14,
against 0.17--0.25 for always predicting the base rate; the expected
calibration error was 0.03; and the residual at contested compositions
($l/k \approx \tfrac12$) lay between $-0.02$ and $+0.01$. The weights,
therefore, were right. The field was not. The four experimental claims,
which had been chosen for near-neutral fields, carried fields of 0.16--0.55
under the experimental scaffold, whereas the calibration set's fields
averaged 0.94. In other words, the blind rule assumed a threshold lower for
the correct side than the experimental claims actually had.

We then asked whether the theory, given the right fields, predicts what we
observed. Re-inserting the four fields into the otherwise unchanged theory
predicted the observed phase: a single wrong-consensus attractor for three
of the claims at every $\alpha$ in the window ($x^* = 0.02$--$0.11$), and
for the fourth a wrong basin extending to $x_0 \approx 0.7$ up to
$\alpha = 0.55$. Rolled out for eight rounds on the realized graphs and
seedings, the same rule reproduced the wrong-consensus rate from low $x_0$
(0.93--1.00 predicted, 0.75--0.99 observed) and the final fractions to an
RMS of 0.09 over the 36 cells. Replacing the realized graphs by the
mean-field independence assumption changed the eight-round prediction by a
further 0.04 (RMS). We call this failure mode field transport: a field
measured on one claim set does not carry over to another, whereas the
message weights do.

A second limit of the blind test is its horizon. The fold of the
mean-field map is a statement about infinite time, whereas the experiment
lasted eight rounds, and near the fold the map relaxes slowly. Indeed, the
blind rule's own eight-round surrogate on the realized graphs fell below
20\% wrong consensus only at $\alpha = 0.63$, not at $\astar = 0.435$; at
$\alpha = 0.45$ it still predicted wrong consensus in 67\% of the low-$x_0$
episodes (Fig.~\ref{fig:collective}a, dashed). The pre-registered
criterion therefore tested the fold through a horizon that displaces it by
about 0.2 in $\alpha$. The two causes can be separated by a counterfactual.
With the blind field, the surrogate fails the 20\% criterion at
$\alpha \le 0.55$ on its own, which is the horizon effect; but at
$\alpha = 1.0$ it predicts wrong consensus in only 0.1\% of the low-$x_0$
episodes, against 75\% observed, which only the field can explain. Thus,
the field is why the prediction failed across the window, and the horizon
is why the criterion could not have resolved the fold near $\astar$ even
with the right field.

\subsection{The discrimination arm favors the normalized rule but does not reject the constant one}

\begin{figure}[t]
\centering
\includegraphics[width=\textwidth]{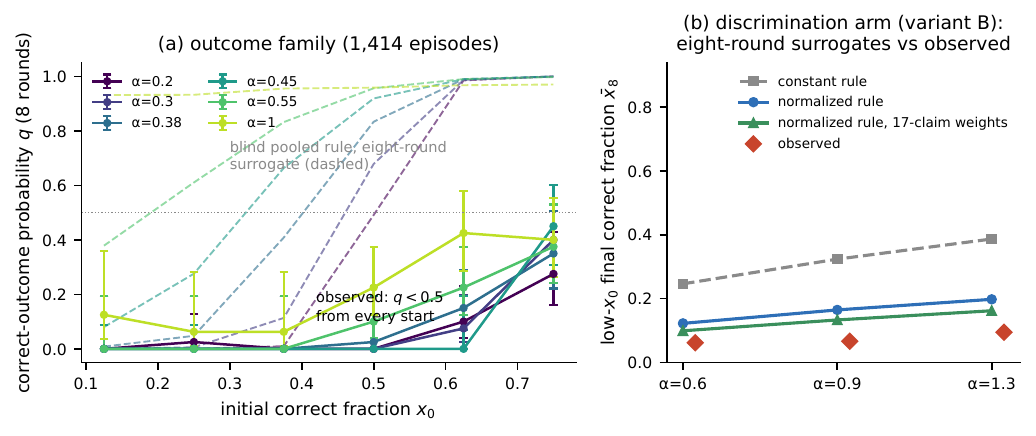}
\caption{\textbf{Outcome probability after eight rounds against the surrogate of the pre-registered rule (arm A), and low-$x_0$ final fractions against surrogates of two coupling rules (arm B).}
(a)~Observed outcome family $q(x_0)$ after eight rounds (solid, Wilson 95\%
intervals) against the eight-round surrogate of the blind pooled rule on the
same graphs and seedings (dashed): the blind prediction is rejected; the
correct-side outcome probability stays below 0.5 from every initial
condition. (b)~Discrimination arm:
observed low-$x_0$ final fractions against the eight-round surrogates of the
normalized (divisive) and constant coupling rules with claim-resolved fields;
the normalized rule lies closer, the constant rule is not rejected.}
\label{fig:collective}
\end{figure}

The discrimination arm survives with a weaker verdict
(Fig.~\ref{fig:collective}b). Under the pooled coefficients, the constant
rule predicted no wrong basin at $\alpha \ge 0.6$ and the normalized rule
predicted absorption up to $\alpha = 1.3$. Wrong-consensus rates of
0.975--1.00 at all three capacities matched the normalized prediction,
which was the pre-registered criterion. With claim-resolved fields,
however, both coupling rules predict wrong-side attractors at these
capacities, and the arm separates them less sharply: the normalized rule
places the low-$x_0$ final fractions at 0.10--0.20 against 0.06--0.10
observed, the constant rule at 0.25--0.39. The data therefore favor the
normalized rule but do not reject the constant one. One claim (1569)
converged wrong-side where both rules predicted correct-side convergence.

Two further checks bound what the collective step adds. Degree-preserving
rewiring ($\rho = 1$ at $\alpha = 0.30$) changed the outcome probability by
at most 0.09 in any cell and by 0.00 in four of six. Because the generated
network is itself uniform given its degrees, this control tests
reproducibility under redrawn sources rather than the sufficiency of
$P(k)$. One residual remains: the collectives ended slightly more correct
than the iterated single-agent rule predicts (final fractions 0.05--0.17
above the surrogate in 11 of the 12 cells at $x_0 \ge 5/8$). In sum, the
collective experiment rejects the blind prediction, verifies each link of
the reduction chain except the transport of the field across claims, and
identifies the claim's field as the quantity that decided the outcome in
the main arm.

\subsection{The field follows the assertion, not the truth: a pre-registered discrimination experiment}

What is the dominant field? It admits three interpretations, which we
separated with predictions registered before data collection: an
assertion-polarity bias (collectives drift toward affirming the claim,
whatever its content), a truth asymmetry (wrong content itself is favored),
and a parametric-prior pull (collectives drift toward the model's stored
world knowledge). Two designs separate them. First, we ran collectives on
three \emph{true} claims (labeled SUPPORTS in CLIMATE-FEVER). Here polarity
and prior predict correct-side convergence, whereas truth asymmetry predicts
wrong-side convergence. Second, we ran collectives on minimally negated
rewrites of the four original misinformation claims. Each rewrite denies
what its original affirms about the same proposition, so that the correct
side switches from denying to affirming; three were made by inserting a
negation and the fourth by replacing ``a non-problem, or even a benefit''
with ``a problem, and not a benefit'' (a contrary rather than a strict
negation, with the quantifier ``often'' retained), and equivalence was
checked by model judgment and by hand. Here polarity predicts that the
drift reverses with the wording, whereas both alternatives predict that it
does not.

\begin{figure}[t]
\centering
\includegraphics[width=\textwidth]{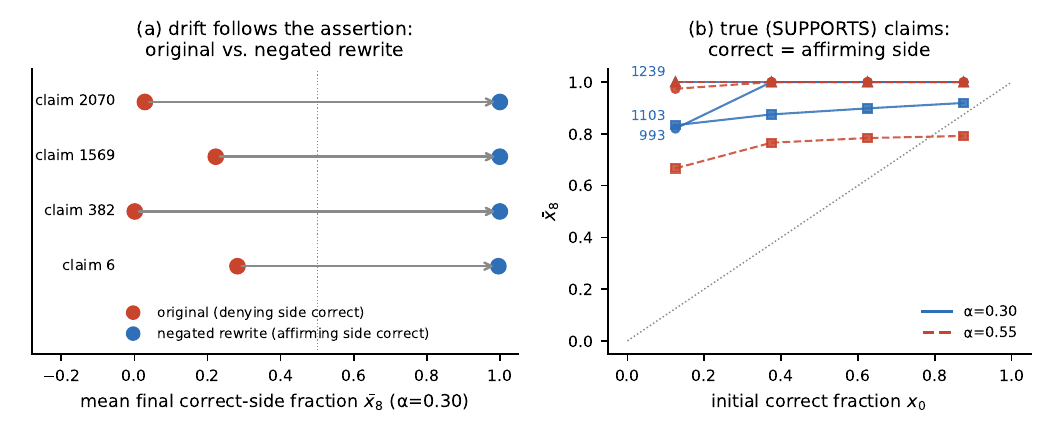}
\caption{\textbf{Final correct-side fractions for original, negated, and true claims.} (a)~Mean final correct-side
fraction for the four original claims (denying side correct) and their
minimally negated rewrites (affirming side correct) under identical
conditions: the content-referenced outcome reverses with the wording.
(b)~True (SUPPORTS) claims converge on the correct = affirming side from all
initial fractions.}
\label{fig:polarity}
\end{figure}

The verdict is polarity (Fig.~\ref{fig:polarity}). On all three true claims
the collectives converged on the correct (affirming) side from every
initial condition, including 12.5\% correct starts (mean final fractions
0.67--1.00). This rejects truth asymmetry. On all four negated rewrites the
collectives converged on the affirming side, which is now the correct side
(correct-side fractions 0.98--1.00), exactly as the originals had converged
on the affirming side when it was the wrong side (correct-side fractions
0.00--0.43). The assertion-referenced outcome is thus the same while the
content-referenced outcome reverses with the wording, which a
parametric-prior pull fixed by content cannot produce. The parametric-prior
account is therefore rejected in the form tested, a content-fixed prior
read from the bare prompt (a quantity that, as
Section~\ref{sec:measurement} notes, does not itself carry over to the full
scaffold), though not in every conceivable form. Note that one rewrite
whose bare-prompt field pointed to the \emph{denying} side ($p = 0.34$)
still converged on the affirming side. This is the one pair in which field
and polarity opposed each other, and the polarity channel overrode the
field rather than merely riding it. Single-agent predictions again matched
the collective outcomes claim by claim in all ten (claim, $\alpha$) cells,
two of them to within the resolution of the seeding grid (SI).

Could the polarity instead be a property of the messages, denying messages
being simply weaker text? Two measurements exclude this sender-side
account. The field itself carries the polarity, with an empty inbox and
before any message is read, and the same 8B-generated message bank read by
llama3.3:70b yields a per-message asymmetry of 1.03 $[0.66, 1.64]$
(Section~\ref{sec:generality}). The asymmetry therefore travels with the
reader, not with the messages. For this model, the collectives follow the
assertion, not the truth. This reframes the social reading of
Section~\ref{sec:results}: the collectives lose to misinformation here
because misinformation is phrased as an affirmation, not because wrong
content is favored. The same reading bears on reports that LLM opinion
networks converge toward scientific ground truth regardless of initial
stance \citep{Chuang2024}. Where the true position is the affirming one, a
polarity channel and a truth bias make identical predictions, and only
designs that reverse the assertion direction, as here, separate them.

\subsection{A second model shows predicted transition points; a third lacks the assertion bias}\label{sec:generality}

\begin{figure}[t]
\centering
\includegraphics[width=\textwidth]{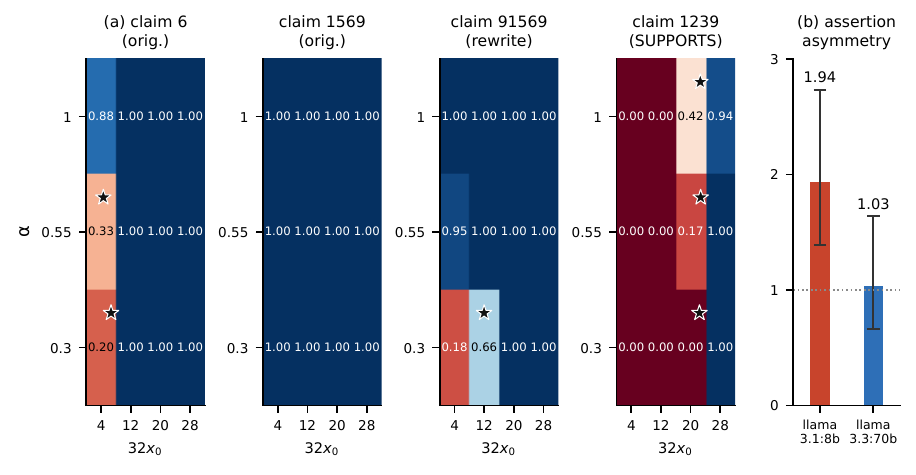}
\caption{\textbf{qwen3:8b collective outcomes against predicted fixed points, and the per-message asymmetry at 70B.} (a)~qwen3:8b
collectives: mean final correct-side fraction per cell; stars mark the
predicted unstable fixed points, which coincide with the observed basin
boundaries for claims 6, 1239, and 91569 at $\alpha = 0.30$; claim 1569 is
predicted and observed monostable correct-side. Cell values are mean final
fractions; intermediate values are splits between the two basins, not
intermediate end states (per-cell outcome counts in the SI). (b)~The
per-message assertion asymmetry on the same eight claims and message bank
(claim-clustered 95\% intervals) is not detected at 70B.}
\label{fig:generality}
\end{figure}

Is the assertion-dominated regime a property of the model or of the
pipeline? To find out, we ran the identical pipeline (claim calibration and
randomized single-shot measurement, then collectives, with fixed-point
predictions computed from the single-shot fits alone) on a second
8-billion-parameter model, qwen3:8b. The outcome was qualitatively
different (Fig.~\ref{fig:generality}). For all twelve qwen3:8b claims the
single-agent data were collected before the corresponding collectives ran,
and the fixed points reported here are computed from those data with the
final estimation procedure (Methods). Their agreement with the collectives
is therefore an out-of-sample test in which no collective outcome enters
the fit, though not a prospective registration; the eight-claim campaign
below adds the prospective element, since its claim-selection rule was
registered before its data existed.

For this model the theory predicts, from the single-agent measurements
alone, claim-dependent bistability: an interior unstable fixed point at
0.14--0.21 for one claim (falling with $\alpha$ and leaving the seeding
grid at $\alpha = 1.0$) and at 0.70--0.71 for another. Both appeared.
Collectives seeded at $x_0 = 0.125$ split between the two basins, while
every start at 0.375 or above converged correctly. For the second claim
the correct-side outcome probability was at most 0.3 from every start at
0.625 or below (mean final fractions 0.00--0.42) and 0.9--1.0 from 0.875,
placing the observed crossings at 0.69--0.75. These are the first observed
interior transition points in this study, and they lie at the computed
positions. For the remaining two claims the pipeline predicts monostable
correct-side convergence in every cell but one (the negated rewrite of
claim 1569 at $\alpha = 0.30$, predicted bistable with transition point
0.37), and the collectives converged correctly in every cell but that one,
where the observed crossing is 0.35 $[0.28, 0.43]$. In sum, all twelve
cells match in class (one to within the resolution of the seeding grid),
five of six predicted transition points fall inside the observed intervals
(the sixth misses by 0.02), and the prediction and observation intervals
overlap in all six. This model is also not assertion-following: one true
claim converged to the \emph{denying} side from all starts below 0.625,
which no polarity account permits.

\begin{figure}[t]
\centering
\includegraphics[width=0.72\textwidth]{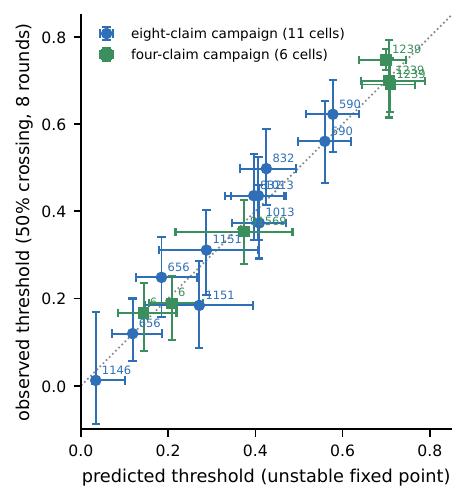}
\caption{\textbf{Calibration of transition-point predictions (qwen3:8b).}
Predicted unstable-fixed-point position (per-claim divisive response
function) against the observed transition point (50\% crossing of the eight-round outcome) for every (claim, $\alpha$) cell observed bistable: circles, the
eight-claim campaign (11 cells); squares, the four-claim campaign (6 cells).
Vertical bars are episode-bootstrap 95\% intervals of the observed transition point, horizontal bars are single-agent-bootstrap 95\% intervals of the
prediction, and the dotted line is identity. Cells predicted and observed
monostable (4 of 16 and 6 of 12) and the one class mismatch (claim 1146 at
$\alpha = 0.30$) are listed in the SI.}
\label{fig:calibration}
\end{figure}

We then took the comparison beyond these four claims with a calibration
campaign. Twenty further claims were calibrated and measured single-shot,
and the same pipeline classified each (claim, $\alpha$) cell as monostable
correct-side, monostable wrong-side, or bistable with a computed transition
point. Eight claims spanning the predicted regimes were selected by a
pre-registered rule applied to the classifications computed at the time of
the campaign (the predictions reported here, recomputed with the estimation
procedure of Methods, leave the selected claims' classes unchanged in 15 of
16 cells), and 256 further episodes were run ($\alpha \in \{0.30, 0.55\}$,
eight initial fractions, zero parse failures). The predicted classification
matched the observed one in 15 of 16 cells (Fig.~\ref{fig:calibration}).
Across the eleven cells observed bistable, all eleven predicted transition
points fall inside the episode-bootstrap intervals, the mean absolute
difference between predicted and observed transition points is 0.04, and
the two are rank-correlated at $r_s = 0.96$ (Spearman; $p = 0.004$ under
permutation of claim labels, claims being the exchangeable unit).
Prediction-side uncertainty, from resampling each claim's single-shot
queries, gives intervals of width 0.08--0.25 on the predicted transition
points, and these overlap the observed intervals in all eleven cells
(Fig.~\ref{fig:calibration}, horizontal bars). The one mismatch is a claim
at $\alpha = 0.30$ whose predicted transition point, 0.09, lies just above
the lowest seeded fraction ($2/32$), from which its collectives
nevertheless converged correctly. At $\alpha = 0.55$ the same claim's
predicted transition point (0.03) and observed crossing (0.01
$[-0.09, 0.17]$) agree, and it is bistable in only 62\% of the resampled
predictions, which marks it as the boundary case. Thus, what was an
existence proof on two claims is a calibration curve on eight: the
single-agent pipeline predicts not only whether a transition point exists
but, to the resolution of the seeding grid, where it sits.

The larger model completes the comparison. We probed llama3.3:70b at the
single-agent level with the same eight claims and message bank. Its
per-message assertion asymmetry is 1.03 $[0.66, 1.64]$, against 1.44
$[1.25, 1.65]$ for llama3.1:8b over the 17 calibration claims and 1.94
$[1.39, 2.73]$ on these same eight (difference 0.90 $[0.00, 1.72]$,
claim-clustered). The bare fields align with content for seven of the eight
claims: with an empty inbox, $P(\mathrm{TRUE}) = 0.000$ for the four false
originals and $1.000$ for three of the four true rewrites, while the
eighth, the negated polar-bear rewrite, is near neutral at
$P(\mathrm{TRUE}) = 0.42$. The polarity channel that governed the
8-billion-parameter collectives is thus not detected, and the larger
model's collective regime (not measured here, a stated limitation) would be
field-dominated on most realistic claims. Note that the larger model
differs from llama3.1:8b in model generation as well as in scale, so the
comparison bounds the channel in the larger model rather than isolating
scale as its cause. In sum, whether a collective has a transition point at
all, where it sits, and what force drives consensus are, across the two
models whose collectives were run, and for the polarity channel in the
third, properties measurable on the individual agent before any collective
is run.

\section{Discussion}\label{sec:discussion}

In the current study, we asked to what extent the reading bound of a
single agent decides where a collective of language-model agents ends up.
We turned the bound into a network with one parameter, the message
capacity $\alpha$, measured the response function of an isolated agent, and
found that the agent behaves as a stochastic binary neuron with divisively
normalized weights. From these measurements alone we predicted the
capacity at which the wrong-consensus basin should disappear, and we
registered the prediction before any collective ran. The prediction
failed. The failure was traced to the threshold that a claim's wording
sets, which did not carry over from the calibration claims to the
experimental ones, and, with each claim's own threshold, the same weights
reproduced the collectives. A pre-registered discrimination experiment then
showed that the threshold follows what a claim asserts rather than whether
it is true, and a second model showed transition points at the positions
the pipeline computed.

The reading bound does not decide a collective's fate on its own. Across
three models it did so together with one more single-agent quantity. The
bound, entering through the degree distribution $P(k)$ of the
crowding-generated network, sets where the transition point sits; the
threshold that a claim's wording sets decides which phase the collective is
in, and whether a transition point exists at all. In llama3.1:8b the
threshold, and the assertion bias behind it, overrode the bound: three of
the four claims sat in a wrong-side monostable phase at every capacity
tested, and 75\%-correct majorities were lost in most episodes. In qwen3:8b
the threshold left claims bistable, and the transition points appeared
where the measured neuron and the crowding network placed them (15 of 16
cells). At 70B the assertion bias was not detected. In every case the
prediction used single-agent measurements only. In every case, too, the
$\alpha$ trend of the wrong-consensus attractor, observed and predicted,
showed that the number of messages each agent was actually shown remained
the operative degree. Where the blind prediction failed, the failure lay
in transporting the field across claim sets, not in the network.

The blind test carries two lessons for prediction in this setting. The
field must be measured on the claims that will be debated: message weights
pooled over a calibration set transfer, the field does not. And the
decision criterion must be computed for the horizon of the experiment. The
fold of the mean-field map is an infinite-time statement, and near the fold
the map relaxes slowly, so an eight-round experiment must be compared with
an eight-round surrogate rather than with the fold itself.

What the collectives verified should be stated in three parts. Verified:
transition points at fixed capacity, crossings in the initial fraction that
appeared at the positions the pipeline computed (qwen3:8b, 15 of 16 cells;
the polarity cells of llama3.1:8b, 10 of 10). Predicted and not observed:
the fold in $\alpha$ itself, the vanishing of the wrong basin as capacity
is reduced, which llama3.1:8b's fields kept out of reach by placing three
of the four claims in the wrong-side monostable phase at every capacity
tested and the fourth in a bistable phase whose transition point lay near
$x_0 \approx 0.7$. Observed as capacity dependence: the movement of the
qwen3:8b transition points with $\alpha$ (claim 6: 0.21 to 0.14 predicted,
0.19 to 0.17 observed, then below the seeding grid at $\alpha = 1.0$), the
same fold approached from the side of the initial fraction. The title's
claim is the first of these: capacity sets where the transition point sits,
and wording sets the threshold and with it which phase the collective is
in.

What $\alpha$ corresponds to outside this experiment differs by system.
Here $\alpha$ is an enforced reading budget: in this protocol the router,
not the model, implements the acceptance kernel, a division of labor the
measurements themselves selected (Section~\ref{sec:measurement}). In
deployed multi-agent systems its counterparts are the quantities that bound
how many messages an agent reads: finite context windows, per-token costs,
and the sampling policies of agent frameworks. In the naming-game
literature the analogous bound is a finite interaction memory
\citep{Flint2026}; in human information sharing, divided attention limits
message visibility in the same direction \citep{HodasLerman2012}. The
model's own capacity enters the theory separately, as the measured load
attenuation $\beta(k)$. Thus, the intake budget sets who is read, and the
attenuation sets how much each reading counts.

Dialogue does not correct error in the field-dominated phase. Majorities of
75\% correct agents ended correct-side in fewer than half of the episodes
at every capacity tested, and the discrimination experiment shows that the
drive is the assertion, not the content. Communication volume and capacity
are therefore no remedy where the threshold set by the wording is adverse.
Accounts that model collective knowledge formation as decentralized
Bayesian inference over exchanged messages \citep{Taniguchi2024} treat
each message as evidence. In the field-dominated phase a message moves its
reader by what it asserts, so the collective aggregates wording rather than
evidence, and convergence to the truth fails under a condition that can be
measured on a single agent. The misinformation-vulnerability reading must
be
stated carefully: small llama collectives lose to misinformation because
misinformation affirms, whereas models without the polarity channel (qwen,
70B) fail or succeed by different routes. Directional conformity has
likewise been reported to collapse majority-vote aggregation in LLM safety
panels under peer pressure \citep{HuQu2026}. The polarity channel measured
here is a claim-level analogue that requires no peer opinions at all, since
it is present in bare single-shot calibration.

The origin of the polarity channel is probed and bounded rather than
resolved. An abliterated (refusal-direction-ablated, \citealp{Arditi2024})
derivative of the same instruct model retains the channel: per-message
asymmetry 1.24 $[1.05, 1.47]$ against 1.44 $[1.25, 1.65]$ for the intact
model, measured in the same batch with the same message bank. Removing
refusal behavior therefore does not remove the assertion bias, and any
reduction is below this measurement's resolution. Whether the channel
exists before instruction tuning could not be measured. The base model does
not follow the forced-choice format (parse rate below 5\% even with a
three-shot format prefix), which places pre-tuning behavior outside any
forced-choice probe; a log-probability readout is the instrument for that
question. The channel is adjacent to, but not identical with, sycophancy as
usually defined, agreement with a user's stated opinion, traced in part to
the preference data used for RLHF \citep{Sharma2023}; here the agreement is
with the claim's own assertion, with no user opinion in the prompt. What
can be said is negative and useful: the bias that decides these
collectives' fate is not carried mainly by the safety-refusal direction.

What is verified here is the reduction chain from single-agent response and
degree statistics to the basin boundaries of \emph{LLM} collectives at
fixed capacity; the fold in capacity itself remains a theoretical
prediction. The suggestion for human groups is nonetheless concrete. The
cognitive limits cited above give human participants a finite effective
intake of the same shape, and the theory's inputs (response threshold, load
attenuation, intake bound) are all behavioral, none tied to a machine
substrate. If deliberating groups sit at finite effective $\alpha$, the
theory would say which of them can recover the truth from a correct
minority and which cannot. Statements about human collectives are,
however, implications rather than measurements. Locating a human group on
the computed $\astar$ curve would need, besides an estimate of a human
$\alpha$ from intake-saturation data, the human counterparts of the
response coefficients (the threshold, the per-message weights, and their
load attenuation), none of which is measured here. A further limitation is
the REFUTES-skew of the calibrated claim set, which restricts all claims to
correct-side/wrong-side language. Identifying genuine truth asymmetry needs
a claim set balanced in polarity at matched difficulty, which this model's
one-sided accuracy on SUPPORTS claims may make impossible. The mitigation
is that the $\astar$ prediction never invoked truth asymmetry: the theory
weights correct-side and wrong-side messages by their measured
coefficients, and the measured inequality $|\beta_T| \ne |\beta_F|$
($\delta < 0$) is, for this REFUTES-skewed set, the assertion asymmetry
seen from the correct side, not a truth asymmetry.

Relative to \citet{El2026}, their graphs were given, ours are generated;
their fits describe the collective data, ours are computed from
single-agent data alone. The two studies probe orthogonal axes of one phase
diagram for the same class of system, binary units with logistic
(Glauber-type) updates. Their transition is thermal: topology fixed, noise
swept, a critical point located by a susceptibility peak, on symmetric
couplings that admit an energy function. Ours is topological: effective
noise fixed (absorbed in the measured $\beta$), degree statistics swept
through $\alpha$, and the transition is a fold, the disappearance of a
metastable basin, a spinodal rather than a critical point. This is why the
collapse of the wrong basin and not a fluctuation peak is the correct
observable. Our graphs are directed, so the dynamics are non-equilibrium
and free-energy arguments do not transfer; the mean-field map does.
Model-dependent collective regimes have now been reported in the
coordination setting as well, in naming-game populations whose collective
bias amplifies, induces, or reverses individual bias with model-specific
size thresholds \citep{Flint2026}, convergent with the model dependence
measured here; and the structured interaction topologies those authors
list as future work are what the crowding process supplies. For
consensus-scaling observations \citep{DeMarzo2024}, $(\log N)/\alpha$
supplies the analytic form for the reported size-limited consensus.

Part of the logistic form is guaranteed by the output mechanism. For a
forced choice between two options, $P(A \mid A\ \text{or}\ B) =
\sigma((z_A - z_B)/T)$ is exactly a logistic function of the logit
difference, and with zero parse failures in every collective campaign the
measured choice probabilities are close to this conditional quantity. What
the mechanism does not guarantee, the empirical content of the reduction,
is that this logit difference is approximately additive in the inbox
composition, one weight per correct-side and per wrong-side message
(Fig.~\ref{fig:neuron}b), as an aggregator of conditionally independent
evidence would produce. Whether the transformer approximates such a
computation, and why, is left open alongside the mechanism of the
attenuation below.

The measured update rule (Fig.~\ref{fig:neuron}a) departs from a classical
Ising spin in exactly one place: the per-message coupling is attenuated by
the total inbox size, $\beta/(1+\gamma k)$, the form known in sensory
neuroscience as divisive normalization \citep{CarandiniHeeger2012}. As in
that literature the law is phenomenological, a behavioral input--output
relation rather than an identified mechanism. Three candidate mechanisms
make separable predictions. Softmax attention, itself a divisive operation,
predicts attenuation governed by token volume rather than message count.
Proportion coding, judging by the fraction $l/k$ of supporting messages as
classically reported for human majority influence, predicts dependence on
$l/k$ alone. Calibration compression of the output probabilities would mimic
attenuation and is quantified by reading log-probabilities directly. A
factorial manipulation of message length and proportional $(l, k)$ scaling
within the Section~\ref{sec:measurement} protocol would separate the three.
We leave this as a stated open question, since no result in this paper
depends on which mechanism generates the measured form.

One residual remains to be explained. The collectives ended slightly more
correct than the iterated single-agent rule predicts, and the residual is
small but systematic (11 of 12 cells, same sign), so it measures a
collective contribution beyond the iterated single-agent rule rather than
noise. One candidate is the evolution of message content across rounds: the
influence of correct-side messages grows by $+0.026$ per round in the
one-step regression (SI), which could reflect either the messages or
repeated exposure. Replaying single-shot queries with round-specific
message banks built from the logged episodes would separate the two, and
the logs permit it.

\section*{Conclusion}

An agent that reads only a few of the others' contributions was the
starting point of this paper. Two single-agent measurements turn that bound
into a prediction: the agent's input-output rule, equivalent to a
stochastic binary neuron with divisively normalized weights, and the
threshold a claim's wording sets for its answer. Placed on a
crowding-generated network, the neuron yields a one-dimensional map whose
fold locates the message capacity at which the wrong-consensus basin
disappears. The blind prediction for one model failed because the threshold
did not carry over between claim sets. With each claim's own threshold the
same weights reproduced the outcomes, and in a second model the predicted
transition points appeared where computed. The practical rule is short:
measure the threshold first, then set the budget. Both are single-agent
measurements, and together they account for most of what the collectives
did, with residuals of 0.05--0.17 in final fraction and one claim outside
both coupling rules remaining. Thus an LLM collective can be assessed
before it runs. Which claims a given model will debate to the wrong side is
a quantity measured here on two models, and how much reading capacity would
let a correct minority win is a quantity the theory computes, verified at
fixed capacity and still a prediction as a fold in capacity itself.

\section*{Methods}

\subsection*{Models, serving, and prompts}

All agents are served with ollama. The primary model is llama3.1:8b
(Q4\_K\_M quantization, temperature 0.7 throughout, seed recorded for every
call); the cross-model campaigns use qwen3:8b (thinking disabled) and
llama3.3:70b (judgments only, message banks fixed to the 8B-generated banks).
The judgment prompt asks for a snap A/B judgment of the claim given the
persona and the inbox messages, with a single-character response; variant B
adds a line stating the agent's current answer. Verbatim prompts, the A/B
counterbalancing, message generation, serving configuration, and the model
variants of the polarity-origin probe are given in SI Section~S11. Every
call's prompt hash, seed, raw response, and parse result are logged to JSONL.

\subsection*{Claims and calibration}

Claims are drawn from CLIMATE-FEVER \citep{Diggelmann2020} (SUPPORTS and
REFUTES only; the claim text is presented alone, without evidence). A
two-stage calibration (24 calls per candidate, refined to 64 for candidates
with $p_q \in [0.35, 0.65]$; A/B assignment counterbalanced) measured
per-claim accuracy $p_q$ and field $h_q = \mathrm{logit}(p_q)$. The collective
experiments use per-claim fields re-estimated under the full experimental
scaffold (Section~\ref{sec:measurement}). The negated rewrites of the four
Stage-2 claims (each rewrite denies what its original affirms about the
same proposition; three by inserting a negation, one by replacing ``a non-problem, or even a
benefit'' with ``a problem, and not a benefit'', a contrary that avoids a
double negative and retains the quantifier ``often'') were checked for
semantic opposition by model judgment
(70--85\% of 20 counterbalanced calls per pair) and approved by hand before
any collective run. Full claim texts, rewrites, and calibration values are in
the SI.

\subsection*{Networks, episodes, and controls}

Each episode generates a fresh directed network: every receiver processes the
other 31 agents in an independent uniformly random order, accepting the next
candidate with probability $e^{-\alpha r}$ after $r$ acceptances. Initial
stances are assigned exogenously to a uniformly random subset of
$\lceil 32 x_0 \rceil$ agents. Each of eight rounds consists of a message
phase (every agent with at least one out-edge generates one two-sentence
message conditioned on its current stance, broadcast to all its receivers)
and a judgment phase (each agent answers the snap-judgement prompt with its
current-round inbox; single sample). The network is fixed within an episode
(regenerated only between episodes), so a receiver reads the same senders in
every round, and their states are correlated through the shared graph even
though every message is regenerated each round from its sender's current
stance; the independence assumption of the mean field of
Section~\ref{sec:theory} is therefore an approximation. Its cost is
quantified in Section~\ref{sec:results} by comparing the eight-round
iteration of the mean-field map with surrogate rollouts of the same rule on
the realized graphs (RMS difference 0.04 in the final fraction). A fidelity check with five-sample majority voting (SI Section~S10) gave
sharper absorption, so the single-sample simplification is not the source
of wrong-side absorption. The rewiring control resamples each receiver's sources uniformly
while preserving its in-degree. Outcomes are classified at $\delta = 0.1$
(correct consensus: $x_8 \ge 0.6$; wrong: $x_8 \le 0.4$), with
$\delta \in \{0.05, 0.2\}$ and undecided-exclusion sensitivity analyses in
the SI.

\subsection*{Cross-model calibration campaign}

Twenty further CLIMATE-FEVER claims, disjoint from all earlier sets, were
calibrated under the full scaffold and measured single-shot with qwen3:8b;
eight were selected by a rule fixed in the pre-registered instructions (at least four predicted bistable with interior transition points, the remainder
monostable; no wrong-side-dominant claim existed among the twenty).
Collectives ran the full $\alpha \in \{0.30, 0.55\} \times x_0$ grid with two
replicates (256 episodes; SI Section~S11).

\subsection*{Statistics}

All confidence intervals on measured coefficients are claim-clustered
pairs-bootstrap percentiles (claims are the clusters because episodes within
a claim share their field; resample counts in SI Section~S11).
Attenuation-form comparison uses five-fold cross-validation with folds split
by claim, alongside AIC; the form parameter is profiled on the full data
($\gamma = 0.9$) and re-profiled inside each training fold, with a held-out
claim's fixed effect set to zero, so the cross-validated likelihood scores
prediction for an unseen claim with an unknown field. The divisive $\astar$
of Section~\ref{sec:measurement} uses $\gamma$ profiled per variant (0.53 for
A, 1.32 for B). Per-$k$ coefficients are identified by a joint logistic fit
over all inbox sizes with claim fixed effects anchored by the empty-inbox
cells, because at a single $k$ the columns $1$, $l$, and $k - l$ are
collinear; per-claim response functions for the collective claims use the
reduced form $c_0 + g(k)[\delta k + \bar\beta(2l - k)]$ ($+\theta s$ for
variant B). Fixed points of the mean-field map are bracketed on a grid in $x$
with bisection refinement and classified by $|F'(x)| < 1$; a phase is
bistable when an unstable interior point separates two stable ones, and
$\astar$ is located by bisection in $\alpha$ on that classification.
Prediction-side intervals resample each claim's single-shot outcomes within
claim and design cell and recompute the fixed points.

\emph{Outcome and transition point.} An episode's outcome is its correct-side
fraction after eight rounds, $x_8$; $q(x_0)$ is the fraction of episodes with
$x_8 \ge 0.6$, a finite-horizon quantity compared with the infinite-time
fixed points only through surrogate rollouts. Observed transition points are the 50\% crossings of a logistic fit of the correct outcome on $x_0$ with a weak
ridge penalty; intervals are episode-bootstrap percentiles (500 replicates).
Rank correlations between predicted and observed transition points are tested by
permuting claim labels, claims being the exchangeable unit. Fitting details,
the construction of Fig.~\ref{fig:neuron}, and the bootstrap retention
counts are in SI Section~S11.

\subsection*{Pre-registration timeline}

``Pre-registered'' means throughout that a prediction or decision rule, with
the criteria for judging it, was written to a file before the corresponding
data were collected, with the file's timestamp and SHA-256 hash recorded in
the project repository; these are local records, not a third-party registry.
Three items were pre-registered: (i)~the $\astar = 0.435 \pm 0.03$ prediction
and the decision criteria of the main collective experiment (2026-08-23);
(ii)~the three-hypothesis prediction table of the polarity discrimination
experiment (2026-08-25); (iii)~the claim-selection rule of the eight-claim
calibration campaign (2026-08-27). The single-agent measurements preceded
the collective experiments. The claim-resolved predictions of
Sections~\ref{sec:results}--\ref{sec:generality} were computed after the
collective experiments from single-agent data alone and were not
prospectively registered; the adaptive second-pass allocation followed rules
fixed in the pre-registered design (SI Section~S11).

\subsection*{Use of generative AI}

The author used Anthropic's Claude for code generation and debugging,
execution of the pre-specified experiment pipelines, drafting and editing of
text, and literature search. The research questions, theory, experimental
designs, pre-registered predictions, and interpretations are the author's;
all AI-assisted output was reviewed and fact-checked, and every reported
number is recomputed from the logged data by the scripts in the public
repository. The language models studied here (llama3.1:8b, qwen3:8b,
llama3.3:70b) are the subject of the research, not writing tools.

\subsection*{Data and code availability}

All episode logs (stance histories, full message texts, graphs, seeds, prompt
hashes), single-shot logs, analysis scripts (scripts 50--57 produce every
identified fit, fixed point, surrogate rollout, and figure reported here), and
notebooks are retained in the project repository, to be released publicly on GitHub
(\url{https://github.com/dockmfgit/AI_agent_crowding}); the public
dataset of \citet{El2026} is not redistributed and is available from its
original source \citep{El2026data}. Layout: \texttt{scripts/} (measurement, engines, analysis),
\texttt{results/} (all tables backing the figures), \texttt{data/}
(raw JSONL and episode JSON), \texttt{notes/} (working records including the
pre-registration files).

\bibliographystyle{plainnat}
\bibliography{refs}

\begin{thebibliography}{19}
\providecommand{\natexlab}[1]{#1}
\providecommand{\url}[1]{\texttt{#1}}
\expandafter\ifx\csname urlstyle\endcsname\relax
  \providecommand{\doi}[1]{doi: #1}\else
  \providecommand{\doi}{doi: \begingroup \urlstyle{rm}\Url}\fi

\bibitem[Arditi et~al.(2024)Arditi, Obeso, Syed, Paleka, Panickssery, Gurnee,
  and Nanda]{Arditi2024}
Andy Arditi, Oscar Obeso, Aaquib Syed, Daniel Paleka, Nina Panickssery, Wes
  Gurnee, and Neel Nanda.
\newblock Refusal in language models is mediated by a single direction.
\newblock In \emph{Advances in Neural Information Processing Systems 37
  (NeurIPS 2024)}, 2024.
\newblock arXiv:2406.11717.

\bibitem[Carandini and Heeger(2012)]{CarandiniHeeger2012}
Matteo Carandini and David~J. Heeger.
\newblock Normalization as a canonical neural computation.
\newblock \emph{Nature Reviews Neuroscience}, 13:\penalty0 51--62, 2012.
\newblock \doi{10.1038/nrn3136}.

\bibitem[Cherry(1953)]{Cherry1953}
E.~Colin Cherry.
\newblock Some experiments on the recognition of speech, with one and with two
  ears.
\newblock \emph{Journal of the Acoustical Society of America}, 25\penalty0
  (5):\penalty0 975--979, 1953.
\newblock \doi{10.1121/1.1907229}.

\bibitem[Chuang et~al.(2024)Chuang, Goyal, Harlalka, Suresh, Hawkins, Yang,
  Shah, Hu, and Rogers]{Chuang2024}
Yun-Shiuan Chuang, Agam Goyal, Nikunj Harlalka, Siddharth Suresh, Robert
  Hawkins, Sijia Yang, Dhavan Shah, Junjie Hu, and Timothy~T. Rogers.
\newblock Simulating opinion dynamics with networks of {LLM}-based agents.
\newblock In \emph{Findings of NAACL 2024}, 2024.

\bibitem[Cowan(2001)]{Cowan2001}
Nelson Cowan.
\newblock The magical number 4 in short-term memory: A reconsideration of
  mental storage capacity.
\newblock \emph{Behavioral and Brain Sciences}, 24\penalty0 (1):\penalty0
  87--114, 2001.
\newblock \doi{10.1017/S0140525X01003922}.

\bibitem[De~Marzo et~al.(2026)De~Marzo, Castellano, and Garcia]{DeMarzo2024}
Giordano De~Marzo, Claudio Castellano, and David Garcia.
\newblock {AI} agents can coordinate via majority-following beyond human scale.
\newblock \emph{Science Advances}, 12:\penalty0 eaea6091, 2026.
\newblock \doi{10.1126/sciadv.aea6091}.
\newblock arXiv:2409.02822.

\bibitem[De~Nobili et~al.(2026)De~Nobili, Iyer, Codello, and
  Burioni]{DeNobili2026}
Cristiano De~Nobili, Vijayasri Iyer, Alessandro Codello, and Raffaella Burioni.
\newblock Microscopic dynamics of consensus formation in multi-agent {LLM}
  naming games.
\newblock arXiv:2608.02178, 2026.

\bibitem[Diggelmann et~al.(2020)Diggelmann, Boyd-Graber, Bulian, Ciaramita, and
  Leippold]{Diggelmann2020}
Thomas Diggelmann, Jordan Boyd-Graber, Jannis Bulian, Massimiliano Ciaramita,
  and Markus Leippold.
\newblock {CLIMATE-FEVER}: A dataset for verification of real-world climate
  claims.
\newblock In \emph{Tackling Climate Change with Machine Learning workshop at
  NeurIPS 2020}, 2020.

\bibitem[El et~al.(2026{\natexlab{a}})El, Paeng, Dinc, Su, Erdogan, Pappu, Ye,
  Zhao, Ganguli, and Zou]{El2026}
Batu El, Jinhee Paeng, Fatih Dinc, Shiye Su, Mete Erdogan, Aneesh Pappu,
  Haotian Ye, Wanjia Zhao, Surya Ganguli, and James Zou.
\newblock {Physics of Agents}: Statistical mechanics predicts collective
  behavior of {AI} agents.
\newblock arXiv:2608.16578, 2026{\natexlab{a}}.

\bibitem[El et~al.(2026{\natexlab{b}})El, Paeng, Dinc, Su, Erdogan, Pappu, Ye,
  Zhao, Ganguli, and Zou]{El2026data}
Batu El, Jinhee Paeng, Fatih Dinc, Shiye Su, Mete Erdogan, Aneesh Pappu,
  Haotian Ye, Wanjia Zhao, Surya Ganguli, and James Zou.
\newblock {Physics of Agents}: data and code [dataset].
\newblock GitHub repository,
  \url{https://github.com/batu-el/physics-of-agents}, 2026{\natexlab{b}}.
\newblock Commit f571742, accessed 2026-08-18; Apache License 2.0.

\bibitem[Flint et~al.(2026)Flint, Aiello, Pastor-Satorras, and
  Baronchelli]{Flint2026}
Ariel Flint, Luca~Maria Aiello, Romualdo Pastor-Satorras, and Andrea
  Baronchelli.
\newblock Group size effects and collective misalignment in {LLM} multi-agent
  systems.
\newblock \emph{Proceedings of the National Academy of Sciences}, 123\penalty0
  (34):\penalty0 e2531697123, 2026.
\newblock \doi{10.1073/pnas.2531697123}.

\bibitem[Flint~Ashery et~al.(2025)Flint~Ashery, Aiello, and
  Baronchelli]{Ashery2025}
Ariel Flint~Ashery, Luca~Maria Aiello, and Andrea Baronchelli.
\newblock Emergent social conventions and collective bias in {LLM} populations.
\newblock \emph{Science Advances}, 11\penalty0 (20):\penalty0 eadu9368, 2025.
\newblock \doi{10.1126/sciadv.adu9368}.

\bibitem[Fukushima(2026)]{SciRepCrowding}
Makoto Fukushima.
\newblock Analytically tractable model of synaptic crowding explains emergent
  small-world structure and network dynamics.
\newblock \emph{Scientific Reports}, 16:\penalty0 11748, 2026.
\newblock \doi{10.1038/s41598-026-47213-2}.

\bibitem[Hodas and Lerman(2012)]{HodasLerman2012}
Nathan~O. Hodas and Kristina Lerman.
\newblock How visibility and divided attention constrain social contagion.
\newblock In \emph{Proc.\ 2012 ASE/IEEE International Conference on Social
  Computing (SocialCom)}, 2012.

\bibitem[Hu and Qu(2026)]{HuQu2026}
Yibo Hu and Jiaming Qu.
\newblock Social pressure breaks majority voting in {LLM} safety panels.
\newblock arXiv:2608.04415, 2026.

\bibitem[Luck and Vogel(1997)]{LuckVogel1997}
Steven~J. Luck and Edward~K. Vogel.
\newblock The capacity of visual working memory for features and conjunctions.
\newblock \emph{Nature}, 390\penalty0 (6657):\penalty0 279--281, 1997.
\newblock \doi{10.1038/36846}.

\bibitem[Sharma et~al.(2024)Sharma, Tong, Korbak, Duvenaud, Askell, Bowman,
  et~al.]{Sharma2023}
Mrinank Sharma, Meg Tong, Tomasz Korbak, David Duvenaud, Amanda Askell,
  Samuel~R. Bowman, et~al.
\newblock Towards understanding sycophancy in language models.
\newblock In \emph{International Conference on Learning Representations (ICLR
  2024)}, 2024.
\newblock arXiv:2310.13548.

\bibitem[Taniguchi et~al.(2025)Taniguchi, Takagi, Otsuka, Hayashi, and
  Hamada]{Taniguchi2024}
Tadahiro Taniguchi, Shiro Takagi, Jun Otsuka, Yusuke Hayashi, and Hiro~Taiyo
  Hamada.
\newblock Collective predictive coding as model of science: Formalizing
  scientific activities towards generative science.
\newblock \emph{Royal Society Open Science}, 12:\penalty0 241678, 2025.
\newblock arXiv:2409.00102.

\bibitem[Treisman(1964)]{Treisman1964}
Anne~M. Treisman.
\newblock Selective attention in man.
\newblock \emph{British Medical Bulletin}, 20\penalty0 (1):\penalty0 12--16,
  1964.
\newblock \doi{10.1093/oxfordjournals.bmb.a070274}.

\end{thebibliography}

\clearpage
\setcounter{section}{0}
\setcounter{figure}{0}
\setcounter{table}{0}
\setcounter{equation}{0}
\renewcommand{\thesection}{S\arabic{section}}
\renewcommand{\thefigure}{S\arabic{figure}}
\renewcommand{\thetable}{S\arabic{table}}
\renewcommand{\theequation}{S\arabic{equation}}
\part*{Supplementary Information}
\noindent Supplementary Information for ``Message capacity and claim wording set the transition points of collective truth-finding in language-model networks''. Section numbers, figures, and tables are prefixed with S; references to Sections 1--5, Figures 1--7, and the Methods without a prefix point to the main text above. In the SI and its tables, ``threshold'' denotes the
transition point in the initial fraction (the unstable fixed point), as in the
main text's earlier terminology.

\section{Glossary of terms used in the main text}

\begin{table}[H]
\caption{Glossary of the terms used in the main text.}
\label{tab:terms}
\small
\begin{tabular}{@{}>{\raggedright\arraybackslash}p{0.27\textwidth}p{0.70\textwidth}@{}}
\toprule
Term & Meaning \\
\midrule
Agent & One LLM instance with a persona, holding a binary stance (TRUE or
FALSE) on a claim; $N = 32$ agents form a collective. \\
Claim; correct side & A short factual statement from CLIMATE-FEVER; the
correct side is the side matching its label (mostly the denying side here,
since most claims used are misinformation). \\
Inbox & The $k$ messages an agent reads in a round, $l$ of which argue for
the correct side. \\
Message capacity $\alpha$ (crowding parameter) & An agent that has already
accepted $r$ sources accepts the next with probability $e^{-\alpha r}$;
larger $\alpha$ means fewer messages read (mean about
$\ln(1+\alpha(N-1))/\alpha$; 6.4 sources at $\alpha = 0.435$). \\
Response function & The probability that an agent answers on the correct
side given its inbox; measured as a logistic function of a weighted sum of
the messages. \\
Field $h$ (threshold) & The agent's threshold for the correct side with
its sign reversed: the log-odds of answering on the correct side with an
empty inbox, set by the claim's wording; positive means the wording alone
lowers the threshold for the correct side. It shifts the logistic response;
the steepness is set by the weights. \\
Per-message weights $\beta_T$, $\beta_F$ & The push of one correct-side and
one wrong-side message on the log-odds; both shrink with inbox size as
$\beta/(1+\gamma k)$ (divisive normalization). \\
$x$, $x_0$, $x_8$ & Fraction of agents on the correct side; at the start;
after the eight rounds. \\
Outcome $q(x_0)$; outcome criterion & Probability that a collective
started at $x_0$ ends on the correct side; the criterion is $x_8 \ge 0.6$
for a correct-side and $x_8 \le 0.4$ for a wrong-side consensus. \\
Mean-field map & An equation giving next round's $x$ from this round's $x$;
its fixed points are the possible consensus states. \\
Basin; transition point; bistable, monostable & The starting fractions that
end at one consensus; the unstable fixed point separating the two basins
(also called the basin boundary, crossing, or tipping point); a collective
with two reachable
consensus states is bistable, with one, monostable. \\
Fold; $\astar$ & The capacity at which the wrong-consensus basin disappears
(a saddle-node bifurcation of the map); above $\astar$ even small correct
minorities win. \\
Variant A, variant B; main arm, discrimination arm & Judgment prompt without
and with the agent's current answer shown; the collective runs made under
each variant. \\
Calibration claims; experimental claims & The 17 claims on which the weights
were measured; the 4 claims on which the main collectives ran. \\
Surrogate rollout & The measured single-agent rule simulated for eight
rounds on the actual graphs and seedings: the finite-time prediction
against which collectives are compared. \\
Field transport & Whether a field measured on one claim set carries over to
another; it did not, whereas the weights did. \\
Assertion bias (polarity channel) & The tendency to move toward affirming a
claim as worded, whatever its truth; present in llama3.1:8b, absent in
qwen3:8b, not detected at 70B. \\
\bottomrule
\end{tabular}
\end{table}

\section{Reanalysis of the public dataset}

\subsection{Endogeneity of the initial fraction}

Variance decomposition of $x_0$ and within/between-question outcome
correlations (random-graph family, $\delta = 0.1$), computed over the 6{,}400
objective communities of El et al.:

\begin{center}
\begin{tabular}{lcccc}
\toprule
model & between-question share of $\mathrm{var}(x_0)$ &
$\mathrm{corr}(x_0,\text{outcome})$ & between & \textbf{within} \\
\midrule
Gemma-3n     & 0.995 & 0.802 & 0.852 & \textbf{0.041} \\
Llama-3-8B   & 0.936 & 0.659 & 0.721 & \textbf{0.162} \\
Qwen3.5-9B   & 0.861 & 0.277 & 0.351 & \textbf{0.132} \\
GPT-4o-mini  & 0.503 & 0.298 & 0.321 & \textbf{0.273} \\
\bottomrule
\end{tabular}
\end{center}

Question-clustered versus i.i.d.\ bootstrap intervals for the fitted
$q(x_0)$ crossing (random family):

\begin{center}
\begin{tabular}{lccccc}
\toprule
model & crossing & i.i.d.\ CI & width & question-clustered CI & ratio \\
\midrule
GPT-4o-mini & 0.494 & $[0.477, 0.513]$ & 0.036 & $[0.404, 0.583]$ & 4.9$\times$ \\
Gemma-3n    & 0.285 & $[0.251, 0.315]$ & 0.065 & $[0.163, 0.419]$ & 3.9$\times$ \\
Qwen3.5-9B  & 0.142 & $[0.059, 0.206]$ & 0.147 & $[-1.049, 0.367]$ & 9.6$\times$ \\
Llama-3-8B  & 0.459 & $[0.441, 0.480]$ & 0.039 & $[0.370, 0.564]$ & 5.0$\times$ \\
\bottomrule
\end{tabular}
\end{center}

Two illustrations of the endogeneity: communities at $x_0 \approx 0.40$ on
different questions run to opposite consensuses with near certainty, while
communities at $x_0 = 0.15$ on an easy question converge correctly in 32 of
32 episodes. These are limitations of the identification strategy, not of
the dataset's quality; the same logs support the message-level measurement
of the next subsection. The degenerate degree distributions noted in the
main text concentrate the random-family mass on $k = 2$--$4$, and the
lattice family has fixed degree.

\subsection{Message-level coefficients and the truth asymmetry}

Question-clustered 95\% intervals for the per-message truth asymmetry $w_T$
(43{,}500 observations per model):

\begin{center}
\begin{tabular}{lcccc}
\toprule
model & $w_T$ & cluster 95\% CI & Ising-coupling estimate & contains 1 \\
\midrule
GPT-4o-mini & 1.055 & $[0.996, 1.114]$ & 1.16 & yes (boundary) \\
Gemma-3n    & 1.179 & $[0.731, 1.628]$ & 1.71 & yes \\
Qwen3.5-9B  & 0.900 & $[0.759, 1.041]$ & 1.33 & yes \\
Llama-3-8B  & 1.144 & $[0.972, 1.316]$ & 1.38 & yes \\
\bottomrule
\end{tabular}
\end{center}

The label attenuation $\lambda_{\mathrm{lab}}$ (per-message weight of
``disagree''-labeled edges relative to ``agree''-labeled ones) is
$0.080 \pm 0.044$ (GPT-4o-mini), $0.555 \pm 0.078$ (Gemma), $0.294 \pm 0.051$
(Qwen), $0.288 \pm 0.041$ (Llama); $\theta$ is nonzero in all four models
($t \ge 3.4$, question-clustered).

\section{Single-agent measurement (llama3.1:8b)}

\subsection{Response coefficients}
\begin{center}
\begin{tabular}{llccr}
\toprule
variant & coefficient & estimate & 95\% CI (claim cluster) & $n_{\mathrm{obs}}$ \\
\midrule
A & $c_0$ & 0.725 & [0.516, 0.915] & 10608 \\
A & h & 0.725 & [0.516, 0.915] & 10608 \\
A & $\beta_T$ & 0.419 & [0.353, 0.486] & 10608 \\
A & $\beta_F$ & -0.568 & [-0.701, -0.446] & 10608 \\
A & $w_T$ & 0.738 & [0.603, 0.899] & 10608 \\
B & $c_0$ & -0.758 & [-0.895, -0.627] & 21216 \\
B & h & 0.196 & [0.062, 0.309] & 21216 \\
B & $\theta$ & 1.909 & [1.723, 2.106] & 21216 \\
B & $\beta_T$ & 0.291 & [0.267, 0.319] & 21216 \\
B & $\beta_F$ & -0.326 & [-0.374, -0.287] & 21216 \\
B & $w_T$ & 0.894 & [0.801, 0.976] & 21216 \\
\bottomrule
\end{tabular}
\end{center}

\subsection{Attenuation-form comparison}
\begin{center}
\begin{tabular}{lrrrr}
\toprule
scheme & parameter & log-lik & AIC & CV log-lik/obs \\
\midrule
load\_decay & 0.14 & -14901.6 & 29845.3 & -0.48256 \\
divisive & 0.9 & -14781.8 & 29605.6 & -0.48276 \\
pos\_decay & 0.25 & -15164.0 & 30370.1 & -0.48871 \\
topk & 5.0 & -15470.0 & 30982.0 & -0.49778 \\
null (additive) & 0.0 & -15807.9 & 31655.8 & -0.50911 \\
\bottomrule
\end{tabular}
\end{center}

\subsection{Identified per-$k$ coefficients}

At a single inbox size $k$ the design columns $1$, $l$, and $k-l$ are
collinear ($l + (k-l) = k$), so a logistic fit at fixed $k$ identifies only
$c_0 + k\beta_F$ and $\beta_T - \beta_F$. The per-$k$ coefficients below are
therefore fitted jointly over all inbox sizes, with one fixed effect per claim
anchored by the empty-inbox cells ($k = 0$), which identifies
$\beta_T(k)$ and $\beta_F(k)$ separately at every measured $k$ (design rank
verified: 29/29 columns for variant A, 30/30 for variant B). Intervals are
claim-clustered bootstrap percentiles (200 resamples).

\begin{center}
\small
\begin{tabular}{rcccc}
\toprule
$k$ & $\beta_T(k)$, variant A & $\beta_F(k)$, variant A & $\beta_T(k)$, variant B & $\beta_F(k)$, variant B \\
\midrule
1  & 2.11 [1.58, 2.96] & $-1.53$ [$-2.03$, $-1.12$] & 1.62 [1.29, 1.94] & $-1.52$ [$-1.80$, $-1.25$] \\
2  & 1.23 [0.98, 1.41] & $-1.21$ [$-1.49$, $-0.94$] & 0.78 [0.60, 0.95] & $-1.10$ [$-1.23$, $-0.93$] \\
3  & 0.80 [0.65, 0.96] & $-1.04$ [$-1.30$, $-0.80$] & 0.61 [0.49, 0.72] & $-0.82$ [$-0.92$, $-0.68$] \\
5  & 0.45 [0.35, 0.55] & $-0.72$ [$-0.88$, $-0.57$] & 0.34 [0.28, 0.41] & $-0.50$ [$-0.57$, $-0.42$] \\
8  & 0.31 [0.26, 0.37] & $-0.51$ [$-0.63$, $-0.41$] & 0.24 [0.20, 0.28] & $-0.34$ [$-0.38$, $-0.29$] \\
12 & 0.23 [0.18, 0.27] & $-0.39$ [$-0.48$, $-0.33$] & 0.18 [0.16, 0.21] & $-0.25$ [$-0.28$, $-0.22$] \\
\bottomrule
\end{tabular}
\end{center}

Mean claim fixed effect under the full scaffold: $0.94$ (variant A; range
$-0.70$ to $3.03$ over the 17 claims), $-0.64$ (variant B), with
$\theta = 2.15$ (variant B).

\begin{figure}[h]
\centering
\includegraphics[width=\textwidth]{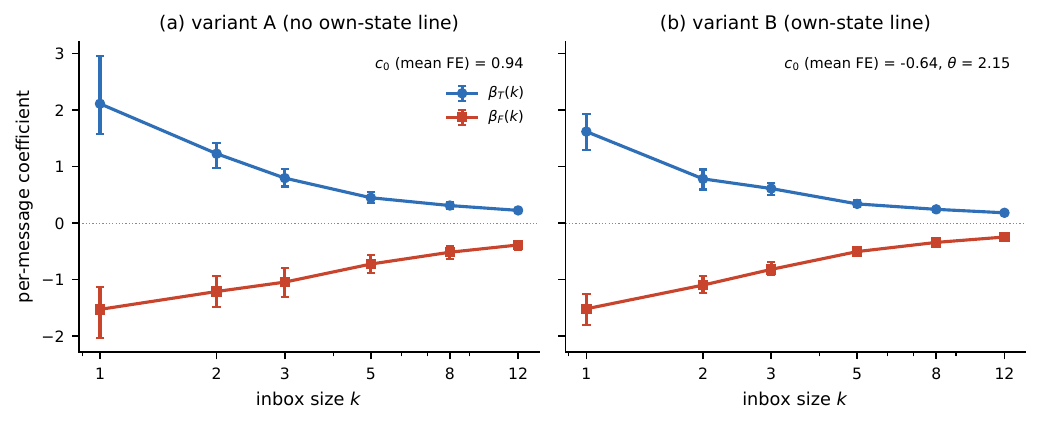}
\caption{\textbf{Identified per-message coefficients versus inbox size
(31{,}824 randomized queries).} $\beta_T(k)$ and $\beta_F(k)$ from the joint
fit with claim fixed effects anchored by the empty-inbox cells, for prompt
variants A (a) and B (b); bars are 95\% claim-clustered bootstrap intervals.}
\label{fig:si_single}
\end{figure}

\section{Reduced-map \texorpdfstring{$\astar$}{alpha*}: validation table}

Fold of the full map $F$ (all $P_\alpha(k)$, exact binomial sums; located by
bisection in $\alpha$) versus the two closure levels of the reduced map $G$,
for seven synthetic coefficient sets and the four measured ones. The fold of
$G$ is solved jointly from $G(x;K) = x$ and $G_x(x;K) = 1$ (residuals below
$10^{-12}$; $G_{xx} > 0$ and $G_K < 0$ at every fold), then converted to
$\astar$ by inverting the exact mean degree $E[K](\alpha)$ or the closed
form~(1) of the main text. Synthetic sets use $\beta_T = -\beta_F = \bar\beta$.

\begin{center}
\small
\resizebox{\textwidth}{!}{\begin{tabular}{lcccc}
\toprule
parameter set & fold of $F$ (exact $\astar$) & $G$-fold, exact $E[K]$ & $G$-fold, closed form & $K^*$ \\
\midrule
$h{=}.10,\ \theta{=}.6,\ \bar\beta{=}.5$ & 0.627 & 0.644 & 0.596 & 4.98 \\
$h{=}.10,\ \theta{=}.6,\ \bar\beta{=}1$ & 1.401 & 1.436 & 1.265 & 2.92 \\
$h{=}.30,\ \theta{=}.6,\ \bar\beta{=}1$ & 1.126 & 1.175 & 1.051 & 3.34 \\
$h{=}.70,\ \theta{=}.6,\ \bar\beta{=}1$ & 0.841 & 0.892 & 0.812 & 4.02 \\
$h{=}.10,\ \theta{=}0,\ \bar\beta{=}1$ & 1.180 & 1.210 & 1.079 & 3.28 \\
$h{=}.10,\ \theta{=}2,\ \bar\beta{=}.5$ & 0.901 & 0.908 & 0.826 & 3.97 \\
$h{=}.30,\ \theta{=}2,\ \bar\beta{=}1$ & 1.545 & 1.592 & 1.392 & 2.72 \\
variant A, measured (constant) & 0.425 & 0.439 & 0.412 & 6.36 \\
variant B, measured (constant) & 0.470 & 0.474 & 0.444 & 6.07 \\
variant A, measured (divisive, $\gamma{=}0.526$) & 0.476 & 0.503 & 0.470 & 5.84 \\
variant B, measured (divisive, $\gamma{=}1.318$) & 1.277 & 1.310 & 1.162 & 3.11 \\
\bottomrule
\end{tabular}
}
\end{center}

Median relative error: 2.7\% (exact-$E[K]$ inversion), 6.7\% (fully closed);
maximum 6.0\% (exact-$E[K]$ inversion; the synthetic set $h = 0.7$,
$\theta = 0.6$, $\bar\beta = 1$) and 9.9\% (fully closed; $h = 0.3$,
$\theta = 2$, $\bar\beta = 1$). The registered variant-A value
$\astar = 0.435$ ($K^* = 6.4$) was obtained with $K^*$ located on a 0.05 grid;
the fold solved to full precision gives 0.439 ($K^* = 6.36$).

Main text Section 3.2 quotes the folds of the full map $F$ (first numeric
column): variant A 0.43 (constant), 0.48 (divisive), 0.46 (identified
per-$k$); variant B 0.47 (constant) and 1.28 (divisive). The registered
prediction $\astar = 0.435$ is the $G$-fold with exact $E[K]$ (second
column).

\emph{Coefficient sets and $\gamma$ by figure.} The constant-coupling sets are
the pooled Stage 1 fits (main text Section 3.2). The divisive $\astar$ values
use $\gamma$ profiled separately per prompt variant ($\gamma_A = 0.526$,
$\gamma_B = 1.318$); Fig.~3 of the main text and the per-claim response
functions of the collective experiments use the pooled profile $\gamma = 0.9$
(Section~S3.2). Fig.~2 of the main text plots the table above.

\emph{Range of the closed form.} $\bar K(\alpha) = \ln(1 + \alpha(N-1))/\alpha$
underestimates the exact mean degree by 2.5\% ($\alpha = 0.2$), 4.0\%
(0.435), 7.0\% (1.0), 11.3\% (2.0) and 68\% ($\alpha = 20$): as
$\alpha \to \infty$ the closed form tends to zero whereas the exact mean tends
to one (the first candidate is always accepted). All $\astar$ values in this
paper lie in $\alpha \le 1.6$, where the underestimate is below 10\%, and the
exact recursion is inverted wherever precision matters.

\section{Main collective experiment}

\subsection{Distribution of final fractions}

The outcome labels of the main text ($x_8 \ge 0.6$ correct-side consensus,
$x_8 \le 0.4$ wrong-side) are operational thresholds. The table gives, per
campaign, the share of episodes that were decided and the share of decided
episodes whose final fraction lies beyond $|x_8 - 0.5| > 0.4$, with the
median end state on each side. Pooled over all 2{,}620 episodes, 66\% end in
$|x_8 - 0.5| \in [0.45, 0.5]$ (unanimity included).
\begin{center}
\small
\resizebox{\textwidth}{!}{\begin{tabular}{lrrrrr}
\toprule
campaign & episodes & decided ($|x_8-0.5|\ge 0.1$) & of which $|x_8-0.5|>0.4$ & median $x_8$, correct side & median $x_8$, wrong side \\
\midrule
main arm A ($\rho=0$) & 1008 & 0.895 & 0.634 & 0.922 & 0.031 \\
main arm A, rewired ($\rho=1$) & 96 & 0.969 & 0.667 & 1.000 & 0.031 \\
discrimination arm B & 300 & 0.917 & 0.633 & 0.875 & 0.062 \\
true claims (A-1) & 288 & 0.976 & 0.758 & 1.000 & --- \\
negated rewrites (A-2) & 192 & 1.000 & 0.995 & 1.000 & --- \\
qwen3:8b, eight claims & 256 & 0.996 & 0.992 & 1.000 & 0.000 \\
qwen3:8b, four claims & 480 & 0.996 & 0.967 & 1.000 & 0.000 \\
\bottomrule
\end{tabular}
}
\end{center}

\subsection{Outcome family (arm A, generated graphs)}
\begin{center}
\small
\begin{tabular}{rrrrccccc}
\toprule
$\alpha$ & $32x_0$ & $n$ & $q$ & Wilson 95\% & undec. & $q_{\delta=.05}$ & $q_{\delta=.2}$ & $q_{\text{excl}}$ \\
\midrule
0.2 & 4 & 16 & 0.000 & [0.000, 0.194] & 0.000 & 0.000 & 0.000 & 0.000 \\
0.2 & 8 & 40 & 0.025 & [0.004, 0.129] & 0.000 & 0.025 & 0.000 & 0.025 \\
0.2 & 12 & 16 & 0.000 & [0.000, 0.194] & 0.000 & 0.000 & 0.000 & 0.000 \\
0.2 & 16 & 16 & 0.000 & [0.000, 0.194] & 0.000 & 0.000 & 0.000 & 0.000 \\
0.2 & 20 & 40 & 0.100 & [0.040, 0.231] & 0.000 & 0.100 & 0.100 & 0.100 \\
0.2 & 24 & 40 & 0.275 & [0.161, 0.428] & 0.000 & 0.275 & 0.275 & 0.275 \\
0.3 & 4 & 40 & 0.000 & [-0.000, 0.088] & 0.025 & 0.000 & 0.000 & 0.000 \\
0.3 & 8 & 16 & 0.000 & [0.000, 0.194] & 0.062 & 0.062 & 0.000 & 0.000 \\
0.3 & 12 & 16 & 0.000 & [0.000, 0.194] & 0.000 & 0.000 & 0.000 & 0.000 \\
0.3 & 16 & 16 & 0.000 & [0.000, 0.194] & 0.125 & 0.000 & 0.000 & 0.000 \\
0.3 & 20 & 40 & 0.075 & [0.026, 0.199] & 0.200 & 0.100 & 0.050 & 0.094 \\
0.3 & 24 & 40 & 0.400 & [0.263, 0.554] & 0.075 & 0.425 & 0.400 & 0.432 \\
0.38 & 4 & 16 & 0.000 & [0.000, 0.194] & 0.000 & 0.000 & 0.000 & 0.000 \\
0.38 & 8 & 16 & 0.000 & [0.000, 0.194] & 0.062 & 0.000 & 0.000 & 0.000 \\
0.38 & 12 & 16 & 0.000 & [0.000, 0.194] & 0.125 & 0.000 & 0.000 & 0.000 \\
0.38 & 16 & 40 & 0.025 & [0.004, 0.129] & 0.075 & 0.025 & 0.000 & 0.027 \\
0.38 & 20 & 40 & 0.150 & [0.071, 0.291] & 0.050 & 0.200 & 0.150 & 0.158 \\
0.38 & 24 & 40 & 0.350 & [0.221, 0.505] & 0.125 & 0.350 & 0.325 & 0.400 \\
0.45 & 4 & 40 & 0.000 & [-0.000, 0.088] & 0.025 & 0.000 & 0.000 & 0.000 \\
0.45 & 8 & 40 & 0.000 & [-0.000, 0.088] & 0.150 & 0.000 & 0.000 & 0.000 \\
0.45 & 12 & 16 & 0.000 & [0.000, 0.194] & 0.125 & 0.000 & 0.000 & 0.000 \\
0.45 & 16 & 16 & 0.000 & [0.000, 0.194] & 0.125 & 0.000 & 0.000 & 0.000 \\
0.45 & 20 & 16 & 0.000 & [0.000, 0.194] & 0.312 & 0.062 & 0.000 & 0.000 \\
0.45 & 24 & 40 & 0.450 & [0.307, 0.602] & 0.075 & 0.475 & 0.450 & 0.486 \\
0.55 & 4 & 16 & 0.000 & [0.000, 0.194] & 0.062 & 0.000 & 0.000 & 0.000 \\
0.55 & 8 & 16 & 0.000 & [0.000, 0.194] & 0.125 & 0.000 & 0.000 & 0.000 \\
0.55 & 12 & 16 & 0.000 & [0.000, 0.194] & 0.125 & 0.000 & 0.000 & 0.000 \\
0.55 & 16 & 40 & 0.100 & [0.040, 0.231] & 0.175 & 0.125 & 0.050 & 0.121 \\
0.55 & 20 & 40 & 0.225 & [0.123, 0.375] & 0.175 & 0.225 & 0.150 & 0.273 \\
0.55 & 24 & 40 & 0.375 & [0.242, 0.530] & 0.075 & 0.400 & 0.300 & 0.405 \\
1 & 4 & 16 & 0.125 & [0.035, 0.360] & 0.188 & 0.188 & 0.125 & 0.154 \\
1 & 8 & 16 & 0.062 & [0.011, 0.283] & 0.125 & 0.125 & 0.000 & 0.071 \\
1 & 12 & 16 & 0.062 & [0.011, 0.283] & 0.312 & 0.125 & 0.000 & 0.091 \\
1 & 16 & 40 & 0.225 & [0.123, 0.375] & 0.275 & 0.300 & 0.150 & 0.310 \\
1 & 20 & 40 & 0.425 & [0.285, 0.578] & 0.125 & 0.450 & 0.300 & 0.486 \\
1 & 24 & 40 & 0.400 & [0.263, 0.554] & 0.275 & 0.525 & 0.325 & 0.552 \\
\bottomrule
\end{tabular}
\end{center}

\subsection{Empirical critical crowding level}
\begin{center}
\begin{tabular}{rrcc}
\toprule
$\alpha$ & $n$ (episodes, $x_0\le 8/32$) & wrong-consensus rate & claim-cluster 95\% CI \\
\midrule
0.2 & 56 & 0.982 & [0.946, 1.000] \\
0.3 & 56 & 0.964 & [0.893, 1.000] \\
0.38 & 32 & 0.969 & [0.906, 1.000] \\
0.45 & 80 & 0.912 & [0.775, 1.000] \\
0.55 & 32 & 0.906 & [0.719, 1.000] \\
1 & 32 & 0.750 & [0.438, 1.000] \\
\bottomrule
\end{tabular}
\end{center}
No crossing of the pre-registered 20\% threshold occurs in the window
($\astar_{\mathrm{emp}}$ undefined; 0 of 2{,}000 claim-clustered bootstrap
resamples cross).

\subsection{Discrimination arm (variant B)}
\begin{center}
\begin{tabular}{rrccc}
\toprule
$\alpha$ & $n$ & wrong rate ($x_0\le 10/32$) & 95\% CI & wrong basin retained \\
\midrule
0.6 & 40 & 0.975 & [0.925, 1.000] & True \\
0.9 & 40 & 1.000 & [1.000, 1.000] & True \\
1.3 & 40 & 0.975 & [0.925, 1.000] & True \\
\bottomrule
\end{tabular}
\end{center}

Observed low-$x_0$ ($x_0 \le 10/32$) mean final fractions against eight-round
surrogate rollouts (200 per episode, same graphs and seedings) of three rules
with claim-resolved fields: the normalized rule with the 17-claim identified
per-$k$ weights, and the per-claim reduced fits with divisive and with
constant coupling:
\begin{center}
\small
\resizebox{\textwidth}{!}{\begin{tabular}{lcccc}
\toprule
$\alpha$ & observed $\bar x_8$ & normalized (17-claim weights) & normalized (per-claim fit) & constant (per-claim fit) \\
\midrule
0.6 & 0.062 & 0.099 & 0.122 & 0.246 \\
0.9 & 0.066 & 0.133 & 0.164 & 0.323 \\
1.3 & 0.095 & 0.162 & 0.198 & 0.386 \\
\bottomrule
\end{tabular}
}
\end{center}
The normalized rules lie closer to the observations than the constant rule
but neither reproduces them exactly; claim 1569 converged wrong-side under
both rules' correct-side predictions (main text Section 4.4).

\subsection{Rewiring control}
\begin{center}
\begin{tabular}{rrrrrr}
\toprule
$32x_0$ & $n_{\rho=0}$ & $q_{\rho=0}$ & $n_{\rho=1}$ & $q_{\rho=1}$ & $\Delta q$ \\
\midrule
4 & 40 & 0.000 & 16 & 0.000 & +0.0000 \\
8 & 16 & 0.000 & 16 & 0.000 & +0.0000 \\
12 & 16 & 0.000 & 16 & 0.000 & +0.0000 \\
16 & 16 & 0.000 & 16 & 0.000 & +0.0000 \\
20 & 40 & 0.075 & 16 & 0.062 & -0.0125 \\
24 & 40 & 0.400 & 16 & 0.312 & -0.0875 \\
\bottomrule
\end{tabular}
\end{center}

\subsection{Yoked one-step calibration}
Every realized transition replayed through the response function with the
17-claim identified per-$k$ weights and each claim's own field:
\begin{center}
\small
\begin{tabular}{lrrrrrr}
\toprule
arm & $\alpha$ & transitions & Brier & Brier (climatology) & ECE & residual at $l/k\in[0.4,0.6]$ \\
\midrule
A & 0.2 & 43,008 & 0.095 & 0.174 & 0.034 & -0.017 \\
A & 0.3 & 67,584 & 0.108 & 0.193 & 0.033 & +0.007 \\
A & 0.38 & 43,008 & 0.120 & 0.212 & 0.030 & +0.003 \\
A & 0.45 & 43,008 & 0.112 & 0.198 & 0.032 & -0.003 \\
A & 0.55 & 43,008 & 0.127 & 0.225 & 0.027 & +0.001 \\
A & 1 & 43,008 & 0.144 & 0.247 & 0.030 & -0.015 \\
B & 0.6 & 25,600 & 0.090 & 0.218 & 0.027 & -0.055 \\
B & 0.9 & 25,600 & 0.099 & 0.225 & 0.027 & -0.046 \\
B & 1.3 & 25,600 & 0.115 & 0.220 & 0.035 & -0.040 \\
\bottomrule
\end{tabular}

\end{center}
At contested compositions ($l/k \in [0.4, 0.6]$; last column) the residual
is within $\pm 0.02$ in arm A, whereas in arm B (own-state line present) a
systematic over-prediction of $0.04$--$0.06$ remains.

\section{Polarity discrimination experiment}

\subsection{SUPPORTS-claim collectives (A-1)}
\begin{center}
\begin{tabular}{llllll}
\toprule
claim & $\alpha$ & $32x_0=4$ & $32x_0=12$ & $32x_0=20$ & $32x_0=28$ \\
\midrule
993 & 0.3 & 0.820 & 1.000 & 1.000 & 1.000 \\
993 & 0.55 & 0.974 & 1.000 & 1.000 & 1.000 \\
1103 & 0.3 & 0.833 & 0.875 & 0.898 & 0.919 \\
1103 & 0.55 & 0.667 & 0.766 & 0.784 & 0.792 \\
1239 & 0.3 & 1.000 & 1.000 & 1.000 & 1.000 \\
1239 & 0.55 & 1.000 & 1.000 & 1.000 & 1.000 \\
\bottomrule
\end{tabular}
\end{center}

\subsection{Negated-rewrite collectives (A-2) versus originals}
\begin{center}
\small
\begin{tabular}{lllllll}
\toprule
rewrite & original & $32x_0=4$ & $32x_0=12$ & $32x_0=20$ & $32x_0=28$ & orig.\ mean $\bar x_8$ ($x_0\le 20/32$) \\
\midrule
90006 & 6 & 0.984 & 1.000 & 1.000 & 1.000 & 0.282 \\
90382 & 382 & 1.000 & 1.000 & 1.000 & 1.000 & 0.001 \\
91569 & 1569 & 1.000 & 1.000 & 1.000 & 1.000 & 0.223 \\
92070 & 2070 & 1.000 & 1.000 & 1.000 & 1.000 & 0.029 \\
\bottomrule
\end{tabular}
\end{center}

\subsection{Claim texts, rewrites, and checks}

Selected SUPPORTS claims (bare-prompt $p_q$ in parentheses): 993 (0.708)
``While there has been a mean rise of a little more than 3mm per year
worldwide since the 1990s, in the last decade, the NOAA Virginia Key tide
gauge just south of Miami Beach has measured a 9mm rise annually.''; 1103
(0.792) ``[ocean acidification was] First referenced in a peer-reviewed study
in Nature in 2003''; 1239 (0.792) ``Through decades of research, it has
become clear that human civilization, roughly 6,000 years old, developed
during an unusually stable period for global sea levels.''

Negated rewrites (model opposition-judgment rate; bare-prompt $p_q$ of the
rewrite, TRUE = correct side):

\begin{center}
\small
\begin{tabular}{p{5.6cm}p{5.6cm}cc}
\toprule
original & rewrite & opp.\ rate & $p_q$ \\
\midrule
The polar bear population has been growing. &
The polar bear population has \emph{not} been growing. & 0.70 & 0.344 \\
Sea level rise is decelerating. &
Sea level rise is \emph{not} decelerating. & 0.85 & 0.938 \\
The amount of energy used to construct solar and wind facilities is greater
than they produce in their working lives. &
\ldots is \emph{not} greater than they produce in their working lives. &
0.70 & 0.938 \\
When life is considered, ocean acidification is often found to be a
non-problem, or even a benefit. &
\ldots is often found to be a \emph{problem, and not a benefit}
(contrary; quantifier ``often'' retained). & 0.85 & 1.000 \\
\bottomrule
\end{tabular}
\end{center}

\subsection{Single-agent predictions for the polarity-experiment cells}

Per-claim divisive response functions (main text Methods) give the predicted
phase of each (claim, $\alpha$) cell of the polarity experiment; all ten cells
are predicted monostable correct-side, and all ten converged correct-side.
In two cells (993 at $\alpha = 0.30$, 1103 at $\alpha = 0.55$) a minority of
episodes seeded at the lowest fraction ended wrong-side, so the fitted 50\%
crossing lies below the seeding grid; these are the two cells described in
the main text as matching to within the resolution of the grid.
\begin{center}
\small
\resizebox{\textwidth}{!}{\begin{tabular}{rrlll}
\toprule
claim & $\alpha$ & predicted (per-claim divisive) & observed & obs.\ $\bar x_8$ by $32x_0$ \\
\midrule
993 & 0.3 & mono-correct & correct from every seeded $x_0$ (crossing below grid) & 0.82 / 1.00 / 1.00 / 1.00 \\
993 & 0.55 & mono-correct & mono-correct & 0.97 / 1.00 / 1.00 / 1.00 \\
1103 & 0.3 & mono-correct & mono-correct & 0.83 / 0.88 / 0.90 / 0.92 \\
1103 & 0.55 & mono-correct & correct from every seeded $x_0$ (crossing below grid) & 0.67 / 0.77 / 0.78 / 0.79 \\
1239 & 0.3 & mono-correct & mono-correct & 1.00 / 1.00 / 1.00 / 1.00 \\
1239 & 0.55 & mono-correct & mono-correct & 1.00 / 1.00 / 1.00 / 1.00 \\
90006 & 0.3 & mono-correct & mono-correct & 0.98 / 1.00 / 1.00 / 1.00 \\
90382 & 0.3 & mono-correct & mono-correct & 1.00 / 1.00 / 1.00 / 1.00 \\
91569 & 0.3 & mono-correct & mono-correct & 1.00 / 1.00 / 1.00 / 1.00 \\
92070 & 0.3 & mono-correct & mono-correct & 1.00 / 1.00 / 1.00 / 1.00 \\
\bottomrule
\end{tabular}
}
\end{center}

\section{Cross-model campaigns}

\subsection{qwen3:8b: predicted fixed points versus observed collectives}

Fixed points of the mean-field map under each claim's divisive response
function (fitted from 336 single-shot queries per claim), with 95\%
single-agent-bootstrap intervals on the unstable point, against the observed
mean final fractions and the fitted 50\% crossing:
\begin{center}
\footnotesize
\resizebox{\textwidth}{!}{\begin{tabular}{rrllll}
\toprule
claim & $\alpha$ & predicted fixed points (s = stable, u = unstable) & pred.\ $x_c$ [95\% CI] & observed $\bar x_8$ ($32x_0 = 4/12/20/28$) & obs.\ crossing [95\% CI] \\
\midrule
6 & 0.3 & 0.0004 (s); 0.2087 (u); 1.0000 (s) & 0.21 [0.16, 0.28] & 0.200 / 1.000 / 1.000 / 1.000 & 0.19 [0.10, 0.25] \\
6 & 0.55 & 0.0010 (s); 0.1440 (u); 1.0000 (s) & 0.14 [0.09, 0.22] & 0.325 / 1.000 / 1.000 / 1.000 & 0.17 [0.08, 0.23] \\
6 & 1 & 1.0000 (s) & --- & 0.884 / 1.000 / 1.000 / 1.000 & below grid \\
1239 & 0.3 & 0.0000 (s); 0.6994 (u); 0.9998 (s) & 0.70 [0.64, 0.75] & 0.000 / 0.000 / 0.000 / 1.000 & 0.75 [0.72, 0.77] \\
1239 & 0.55 & 0.0000 (s); 0.7094 (u); 0.9997 (s) & 0.71 [0.64, 0.77] & 0.000 / 0.000 / 0.166 / 1.000 & 0.69 [0.63, 0.75] \\
1239 & 1 & 0.0000 (s); 0.7067 (u); 0.9995 (s) & 0.71 [0.64, 0.79] & 0.000 / 0.000 / 0.422 / 0.944 & 0.70 [0.61, 0.79] \\
1569 & 0.3 & 1.0000 (s) & --- & 1.000 / 1.000 / 1.000 / 1.000 & --- \\
1569 & 0.55 & 1.0000 (s) & --- & 1.000 / 1.000 / 1.000 / 1.000 & --- \\
1569 & 1 & 1.0000 (s) & --- & 1.000 / 1.000 / 1.000 / 1.000 & --- \\
91569 & 0.3 & 0.0195 (s); 0.3738 (u); 0.9940 (s) & 0.37 [0.22, 0.49] & 0.178 / 0.656 / 1.000 / 1.000 & 0.35 [0.28, 0.43] \\
91569 & 0.55 & 0.9958 (s) & --- & 0.953 / 1.000 / 1.000 / 1.000 & --- \\
91569 & 1 & 0.9973 (s) & --- & 1.000 / 1.000 / 1.000 / 1.000 & --- \\
\bottomrule
\end{tabular}
}
\end{center}

Per-cell outcome counts (correct/wrong/undecided of 10 episodes) for the same
campaign; intermediate mean final fractions in Fig.~6 of the main text are
splits between the two basins:
\begin{center}
\small
\begin{tabular}{rrcccc}
\toprule
claim & $\alpha$ & $32x_0=4$ & $32x_0=12$ & $32x_0=20$ & $32x_0=28$ \\
\midrule
6 & 0.3 & 2/8/0 & 10/0/0 & 10/0/0 & 10/0/0 \\
6 & 0.55 & 3/7/0 & 10/0/0 & 10/0/0 & 10/0/0 \\
6 & 1 & 9/1/0 & 10/0/0 & 10/0/0 & 10/0/0 \\
1239 & 0.3 & 0/10/0 & 0/10/0 & 0/10/0 & 10/0/0 \\
1239 & 0.55 & 0/10/0 & 0/10/0 & 2/8/0 & 10/0/0 \\
1239 & 1 & 0/10/0 & 0/10/0 & 3/7/0 & 9/0/1 \\
1569 & 0.3 & 10/0/0 & 10/0/0 & 10/0/0 & 10/0/0 \\
1569 & 0.55 & 10/0/0 & 10/0/0 & 10/0/0 & 10/0/0 \\
1569 & 1 & 10/0/0 & 10/0/0 & 10/0/0 & 10/0/0 \\
91569 & 0.3 & 0/9/1 & 6/4/0 & 10/0/0 & 10/0/0 \\
91569 & 0.55 & 10/0/0 & 10/0/0 & 10/0/0 & 10/0/0 \\
91569 & 1 & 10/0/0 & 10/0/0 & 10/0/0 & 10/0/0 \\
\bottomrule
\end{tabular}

\end{center}

qwen3:8b calibration of the 120 candidate claims found 4 SUPPORTS and 3
REFUTES claims with $|h_q| < 0.4$; a polarity-balanced set of 12 could not be
formed, so genuine truth asymmetry remains unidentified in this model as
well.

\subsection{llama3.3:70b single-agent probe}
\begin{center}
\begin{tabular}{rrrrrr}
\toprule
claim & $c_0$ & $\beta_T$ & $\beta_F$ & $n$ & $P(\text{TRUE})$ at $k=0$ \\
\midrule
6 & 1.987 & 0.636 & -0.257 & 132 & 0.000 \\
382 & 3.920 & 16.536 & -0.500 & 132 & 0.000 \\
1569 & 36.547 & 5.377 & -4.482 & 132 & 0.000 \\
2070 & 3.455 & 1.375 & -0.554 & 132 & 0.000 \\
90006 & 0.378 & 0.487 & -0.494 & 132 & 0.417 \\
90382 & 1.739 & 1.407 & -0.748 & 132 & 1.000 \\
91569 & 1.713 & 1.156 & -0.572 & 132 & 1.000 \\
92070 & 3.910 & 0.203 & 0.203 & 132 & 1.000 \\
\bottomrule
\end{tabular}
\end{center}
Polarity-coded regression across all eight claims (claim fixed effects):
$b(\text{TRUE-side message}) = +0.583$, $b(\text{FALSE-side}) = -0.563$,
ratio 1.03 with claim-clustered 95\% interval $[0.66, 1.64]$ (2{,}000
resamples of the eight claims; weak ridge $10^{-3}$ against separation). On
the same eight claims and message bank llama3.1:8b gives 1.94 $[1.39, 2.73]$
(over the 17 calibration claims: 1.44 $[1.25, 1.65]$); the 8b$-$70B
difference is 0.90 $[0.00, 1.72]$ and the ratio of ratios 1.87
$[1.00, 3.12]$, so the reduction is at the edge of resolution with eight
claims. Per-claim coefficients with $|c_0| > 5$ in the table above reflect
separation (accuracy 0 or 1 at $k = 0$) and are reported for completeness.
Empty-inbox fields (last column, observed rates over the empty-inbox
queries of each claim): the four false originals give
$P(\mathrm{TRUE}) = 0.000$ (every answer on the denying and correct side); of
the four true rewrites, three give 1.000 and the negated polar-bear rewrite
(90006) 0.42, the one near-neutral claim among the eight.
Per-$k$ fits (truth-coded, pooled across claims):
$\beta_T(4) = 2.41 \to \beta_T(8) = 1.35$. The positive wrong-side
coefficient at $k = 4$ ($+0.61$) --- wrong-side messages slightly
\emph{increasing} correct-side probability --- is recorded as an observation
only.

\section{Stage 2c: eight-claim calibration campaign and polarity-origin probe}

\subsection{Calibration campaign: all 16 cells}

Predicted classifications and thresholds come from each claim's divisive
response function fitted to its single-shot queries (Methods); the eight
claims were selected from twenty candidates by the pre-registered rule.
Prediction intervals are single-agent bootstrap percentiles (200 resamples
of each claim's queries within design cells) and ``frac.\ bistable'' is the
fraction of resamples classified bistable. Observed crossings are 50\% points
of ridge-penalized logistic fits of episode outcomes against $x_0$, with
episode-bootstrap 95\% intervals;
$n_{\mathrm{c}}/n_{\mathrm{w}}/n_{\mathrm{u}}$ are correct/wrong/undecided
episode counts of 16 per cell.

\begin{table}[h]
\centering\footnotesize
\resizebox{\textwidth}{!}{\begin{tabular}{rrlllll}
\toprule
claim & $\alpha$ & predicted & pred.\ $x_c$ [95\% CI] (frac.\ bistable) & observed & obs.\ crossing [95\% CI] & $n_{\mathrm{c}}/n_{\mathrm{w}}/n_{\mathrm{u}}$ \\
\midrule
97 & 0.3 & mono-correct & --- & mono-correct & --- & 16/0/0 \\
97 & 0.55 & mono-correct & --- & mono-correct & --- & 16/0/0 \\
184 & 0.3 & mono-correct & --- & mono-correct & --- & 16/0/0 \\
184 & 0.55 & mono-correct & --- & mono-correct & --- & 16/0/0 \\
590 & 0.3 & bistable & 0.559 [0.497, 0.619] (1.00) & bistable & 0.560 [0.465, 0.653] & 7/9/0 \\
590 & 0.55 & bistable & 0.578 [0.516, 0.637] (1.00) & bistable & 0.622 [0.536, 0.701] & 6/10/0 \\
656 & 0.3 & bistable & 0.185 [0.125, 0.267] (1.00) & bistable & 0.248 [0.157, 0.341] & 12/4/0 \\
656 & 0.55 & bistable & 0.119 [0.072, 0.187] (1.00) & bistable & 0.120 [0.057, 0.200] & 14/2/0 \\
832 & 0.3 & bistable & 0.425 [0.364, 0.494] (1.00) & bistable & 0.498 [0.413, 0.589] & 8/8/0 \\
832 & 0.55 & bistable & 0.396 [0.331, 0.466] (1.00) & bistable & 0.436 [0.334, 0.531] & 9/7/0 \\
1013 & 0.3 & bistable & 0.407 [0.344, 0.470] (1.00) & bistable & 0.436 [0.335, 0.525] & 9/7/0 \\
1013 & 0.55 & bistable & 0.409 [0.346, 0.471] (1.00) & bistable & 0.373 [0.292, 0.460] & 10/6/0 \\
1146 & 0.3 & bistable & 0.094 [0.046, 0.157] (0.98) & mono-correct & --- & 16/0/0 \\
1146 & 0.55 & bistable & 0.034 [0.025, 0.101] (0.62) & bistable & 0.012 [-0.087, 0.169] & 15/1/0 \\
1151 & 0.3 & bistable & 0.271 [0.147, 0.395] (0.96) & bistable & 0.184 [0.085, 0.285] & 13/3/0 \\
1151 & 0.55 & bistable & 0.288 [0.178, 0.407] (0.95) & bistable & 0.311 [0.208, 0.404] & 11/4/1 \\
\bottomrule
\end{tabular}
}
\caption{All 16 cells of the eight-claim calibration campaign (qwen3:8b,
256 episodes, zero parse failures). Fifteen cells match in class; the
mismatch is claim 1146 at $\alpha = 0.30$, predicted bistable with threshold
0.094 just above the lowest seeded fraction (2/32), observed monostable
correct. All eleven predicted thresholds of the cells observed bistable lie
inside the observed intervals (mean absolute difference 0.04; Spearman
$r_s = 0.96$, claim-permutation $p = 0.004$).}
\end{table}

\subsection{Polarity-origin probe (D-3)}

Judgment-side single-shot measurements on three llama3.1-8b variants with the
fixed 8B message bank and a three-shot answer-format prefix applied
identically to all variants. The base model does not follow the forced-choice
format (parse rate 1.5\% in the bare pilot, 4.9\% under the prefix), so its row is
reported for completeness and carries no statistical weight.

\begin{table}[h]
\centering\small
\resizebox{\textwidth}{!}{%
\begin{tabular}{lllll}
\toprule
variant & tag (digest) & parsed & $|b_T|/|b_F|$ & 95\% CI \\
\midrule
instruct (control) & llama3.1:8b (46e0c10c039e) & 5{,}984/5{,}984 & 1.437 & [1.250, 1.651] \\
abliterated & mannix/llama3.1-8b-abliterated (f7d729db9832, Q4\_0) & 5{,}984/5{,}984 & 1.243 & [1.051, 1.471] \\
base & llama3.1:8b-text-q4\_K\_M (6f98b5a6e4b7) & 294/5{,}984 & (2.10) & not evaluable \\
\bottomrule
\end{tabular}}
\caption{Per-message assertion asymmetry (claim-fixed-effect logistic
regression; claim-clustered bootstrap intervals). The instruct control
reproduces the Stage-1 value 1.44 under the added format prefix. Bare
affirmation indices $S$ for the four original/negated pairs: instruct
0.92--1.52 (mean 1.22), abliterated 1.13--1.42 (mean 1.24); base cells parsed
2--6 of 64 and are not evaluable.}
\end{table}

\section{Residual and consistency analyses}

\subsection{Message statistics by side}
\begin{center}
\begin{tabular}{rlrrr}
\toprule
claim & side & messages & words/msg & hedges/msg \\
\midrule
6 & correct & 28258 & 60.7 & 0.058 \\
6 & wrong & 35998 & 59.4 & 0.419 \\
382 & correct & 16713 & 63.5 & 0.035 \\
382 & wrong & 47591 & 62.9 & 0.116 \\
1569 & correct & 28115 & 58.8 & 0.084 \\
1569 & wrong & 36093 & 59.6 & 0.314 \\
2070 & correct & 14105 & 64.5 & 0.081 \\
2070 & wrong & 50119 & 59.8 & 1.494 \\
\bottomrule
\end{tabular}
\end{center}

\subsection{Recorded-message persuasiveness probes}
\begin{center}
\begin{tabular}{rllcr}
\toprule
claim & side & message source & $P$(correct) & $n$ \\
\midrule
6 & correct & bank & 1.000 & 60 \\
6 & correct & episode & 1.000 & 60 \\
6 & wrong & bank & 0.100 & 60 \\
6 & wrong & episode & 0.167 & 60 \\
382 & correct & bank & 1.000 & 60 \\
382 & correct & episode & 1.000 & 60 \\
382 & wrong & bank & 0.017 & 60 \\
382 & wrong & episode & 0.000 & 60 \\
1569 & correct & bank & 1.000 & 60 \\
1569 & correct & episode & 1.000 & 60 \\
1569 & wrong & bank & 0.100 & 60 \\
1569 & wrong & episode & 0.050 & 60 \\
2070 & correct & bank & 1.000 & 60 \\
2070 & correct & episode & 1.000 & 60 \\
2070 & wrong & bank & 0.050 & 60 \\
2070 & wrong & episode & 0.167 & 60 \\
\bottomrule
\end{tabular}
\end{center}
Episode-generated and bank messages are indistinguishable in persuasive
force; no message-quality contribution to the high-$x_0$ residual is detected.

\subsection{Repeated exposure and finite time}

One-step regression with round interactions gives
$\beta_T \times (t-1) = +0.026$ (SE 0.006) and $\beta_F \times (t-1) =
-0.004$ (SE 0.003): repeated exposure does not attenuate message influence.
Finite-time check on high-$x_0$ episodes:
\begin{center}
\small
\begin{tabular}{rrrrrr}
\toprule
$\alpha$ & $32x_0$ & $n$ & mean $x_8-x_7$ & frac.\ interior & frac.\ still moving \\
\midrule
0.2 & 20 & 40 & -0.008 & 0.300 & 0.275 \\
0.2 & 24 & 40 & -0.009 & 0.225 & 0.150 \\
0.3 & 20 & 40 & -0.012 & 0.450 & 0.250 \\
0.3 & 24 & 40 & +0.001 & 0.275 & 0.325 \\
0.38 & 20 & 40 & +0.001 & 0.450 & 0.375 \\
0.38 & 24 & 40 & -0.028 & 0.375 & 0.350 \\
0.45 & 20 & 16 & -0.020 & 0.438 & 0.375 \\
0.45 & 24 & 40 & -0.008 & 0.275 & 0.375 \\
0.55 & 20 & 40 & -0.009 & 0.525 & 0.375 \\
0.55 & 24 & 40 & -0.002 & 0.450 & 0.225 \\
1 & 20 & 40 & -0.002 & 0.800 & 0.575 \\
1 & 24 & 40 & -0.027 & 0.725 & 0.550 \\
\bottomrule
\end{tabular}
\end{center}

\subsection{Label residual}

Correlation between an episode's share of TF label assignments and its final
state: $r = 0.194$ ($n = 1{,}008$ episodes) --- an order of magnitude below
the experimental effects ($\Delta \bar x_8 \approx 1$), but nonzero;
per-call randomization protects pooled estimates.

\subsection{Flip statistics and cascades}

Refractory regression: the coefficient of ``flipped at $t-1$'' on flipping at
$t$ is $+0.98$ (claim-cluster SE 0.13) --- flips concentrate in particular
agents (anti-refractory). Branching ratios:
\begin{center}
\begin{tabular}{rrr}
\toprule
$\alpha$ & flip events & $\sigma$ (mean downstream flips) \\
\midrule
0.2 & 9048 & 3.546 \\
0.3 & 10129 & 3.102 \\
0.38 & 11452 & 2.771 \\
0.45 & 10473 & 2.476 \\
0.55 & 12301 & 2.312 \\
1 & 14625 & 1.776 \\
\bottomrule
\end{tabular}
\end{center}
Cascade-size distribution in the asynchronous public dataset (GPT-4o-mini,
320 runs; cascades linked by graph adjacency within three events):
\begin{center}
\begin{tabular}{rr}
\toprule
cascade size & count \\
\midrule
1 & 9341 \\
2 & 668 \\
3 & 130 \\
4 & 28 \\
5 & 9 \\
6 & 1 \\
7 & 1 \\
\bottomrule
\end{tabular}
\end{center}
91.8\% of cascades have size 1; the tail decays geometrically (deeply
subcritical).

\section{Fidelity check}

Ten episodes at $\alpha = 0.30$, $x_0 = 16/32$ re-run with five-sample
majority voting: wrong-consensus rate 1.00, undecided 0.00, mean
$|x_8 - 0.5| = 0.44$, versus 0.875, 0.125, 0.38 under single-sample
judgments --- absorption is sharper, not weaker, under the original
five-sample protocol.

\section{Extended Methods}\label{si:methods}

This section reproduces in full the Methods subsections that the main text
summarizes: models and serving, verbatim prompts, the cross-model
calibration campaign, the complete statistical procedures, and the
pre-registration statement.

\subsection{Models and serving}

All agents are served with ollama. The primary model is llama3.1:8b
(Q4\_K\_M quantization, temperature 0.7 throughout, seed recorded for every
call). The cross-model campaign uses qwen3:8b (tag 500a1f067a9f, thinking
disabled via \texttt{think:false}) and llama3.3:70b (Q4\_K\_M; judgments
only, message banks fixed to the 8B-generated banks). Single-shot campaigns
ran against a user-space ollama instance with \texttt{OLLAMA\_NUM\_PARALLEL=16}
(the system instance saturates at 0.62 generations/s; the parallel instance
reaches 3.5 generations/s); the 70B model does not load under 16-slot KV
allocation and was served by the serial system instance. The polarity-origin probe (Discussion) used two further variants, judgments
only: llama3.1:8b-text-q4\_K\_M (the base model) and an abliterated derivative
of the instruct model (mannix/llama3.1-8b-abliterated, Q4\_0), with a
three-shot answer-format prefix applied identically to all variants,
including the re-measured instruct control. Every call's prompt
hash, seed, raw response, and parse result are logged to JSONL.

\subsection{Prompts}

The judgment prompt (verbatim; variant A) is:
\begin{quote}\footnotesize\ttfamily
\# Your Expertise\\
\{persona\}\\[2pt]
\# Messages\\
You received the following messages from fellow panelists:\\
- \{message 1\} \ldots\\[2pt]
\# Problem\\
You are one member of an expert panel judging the following claim.\\
Claim: "\{claim\}"\\
Options:\\
A) \{TRUE or FALSE\}\\
B) \{FALSE or TRUE\}\\
\# Instruction\\
Give your direct answer on which option is correct.\\
Do NOT work through the problem, show any steps, compute, or\\
explain --- just commit to a snap judgement.\\
\#\# Response Format\\
Respond with a single character, A or B, and nothing else.
\end{quote}
Variant B inserts, before \texttt{\# Problem}: ``\texttt{\# Your Current
Answer / Your current answer is that the claim is \{TRUE|FALSE\}.}'' The A/B
$\leftrightarrow$ TRUE/FALSE assignment is randomized per call and
counterbalanced within cells; message presentation order is randomized and
recorded. The message-generation prompt asks the sender, given its persona,
the claim, and its current stance, to ``write a brief message to a fellow
panelist stating that you believe the claim is \{stance\} and giving the
single strongest reason from your reasoning,'' with a two-sentence response
format; stance is expressed in the message body (TRUE/FALSE), never as an
option letter, because option labels are randomized per receiver. Personas are
one-line expertise statements drawn from a fixed pool of 12. Prompt texts are
extracted verbatim from the released scripts.

\subsection{Cross-model calibration campaign}

Twenty further CLIMATE-FEVER claims (ten SUPPORTS, ten REFUTES, disjoint from
all earlier sets) were calibrated under the full scaffold and measured
single-shot with qwen3:8b (24 calls per candidate, 64 in the boundary band;
per-claim response-function fits from 6{,}720 further queries). The resulting
fixed-point classifications and thresholds for $\alpha \in \{0.30, 0.55\}$
were computed at the time of the campaign, and eight claims were selected by
a rule fixed in the pre-registered instructions (at least four predicted bistable with interior
thresholds, the remainder monostable, filled from the bistable pool where a
category was empty --- no wrong-side-dominant claim existed among the twenty
candidates). Collectives ran the full grid $\alpha \in \{0.30, 0.55\} \times
x_0 \in \{2, 6, \ldots, 30\}/32$ with two replicates (256 episodes, protocol
identical to the main cross-model campaign). Observed thresholds are defined
under Statistics (main text) and in the next subsection.

\subsection{Statistics: full procedures}

All confidence intervals on measured coefficients are claim-clustered pairs
bootstrap percentiles (1{,}000--2{,}000 resamples for the pooled
coefficients, 200 for the identified per-$k$ coefficients and for the
prediction-side intervals, 2{,}000 for the polarity ratios; claims are the
clusters because episodes within a claim share its field). Attenuation-form comparison
uses five-fold cross-validation with folds split by claim, alongside AIC; the
form parameter ($\gamma$ for divisive normalization, the rate for load decay)
is profiled on the full data ($\gamma = 0.9$) and re-profiled inside each
training fold, and a held-out claim's fixed effect, which cannot be estimated
from the training folds, is set to zero, so the cross-validated likelihood
scores prediction for an unseen claim with an unknown field. The divisive
$\astar$ of Section~3 of the main text uses $\gamma$ profiled separately
per variant (0.53 for A, 1.32 for B); Fig.~3 of the main text uses the pooled
$\gamma = 0.9$. Per-$k$ coefficients are identified by a joint logistic fit
over all inbox sizes with claim fixed effects anchored by the empty-inbox
cells: at a single $k$ the columns $1$, $l$, and $k - l$ are collinear, so
separate per-$k$ fits identify only $c_0 + k\beta_F$ and $\beta_T - \beta_F$.
Per-claim response functions for the collective claims use the reduced form
$c_0 + g(k)[\delta k + \bar\beta(2l - k)]$ ($+\theta s$ for variant B), with
a weak ridge penalty ($10^{-3}$) where $k$ takes three values. For Fig.~3 of the main text, one logistic fit per prompt variant
with claim fixed effects and message weights $\beta_T g(k)$, $\beta_F g(k)$
(divisive, $\gamma = 0.9$ fixed, or $g \equiv 1$) assigns each query a
predicted probability; a design cell's net input $u$ is the logit of its mean
predicted probability, and its observed correct-side rate is plotted against
$u$ with a Wilson 95\% interval. Fixed points of the mean-field map are
bracketed on an endpoint-inclusive grid of 4{,}001 points in $x$ with
bisection refinement and classified by $|F'(x)| < 1$; a phase is bistable
when an unstable interior point separates two stable ones, and $\astar$ is
located by bisection in $\alpha$ on that classification, with the fold of $G$
solved jointly from $G = x$ and $G_x = 1$. Prediction-side intervals resample
each claim's single-shot outcomes within that claim and within each design
cell (200 resamples; a within-claim bootstrap, not the claim-clustered
bootstrap used for pooled coefficients) and recompute the fixed points. Reanalysis statistics on the public dataset use
question-clustered bootstrap and CR0 sandwich errors.

\emph{Outcome and threshold.} An episode's outcome is its correct-side
fraction after eight rounds, $x_8$; $q(x_0)$ is the fraction of episodes with
$x_8 \ge 0.6$, a finite-horizon quantity that we compare with the
infinite-time fixed points only through surrogate rollouts. Observed
thresholds are the 50\% crossings of a logistic fit of the correct outcome on
$x_0$ with a weak ridge penalty (0.002) so that separated cells have a
finite estimate; the slope is not constrained, a fit with a non-positive slope
would be discarded, and none occurred in the reported cells. Intervals are
episode-bootstrap percentiles (500 replicates); replicates without a finite
crossing in $(-1, 2)$ are dropped, which retained all replicates in 12 of the
17 cells observed bistable, 97--99\% in three, and 89\% and 65\% in the two
cells whose crossing lies nearest the lowest seeded fraction (claims 656 and
1146 at $\alpha = 0.55$). Rank correlations between predicted and observed thresholds are
tested by permuting claim labels, claims being the exchangeable unit.

\subsection{Pre-registration timeline: full statement}

Throughout this paper, ``pre-registered'' means that a prediction or a
decision rule, together with the criteria for judging it, was written to a
file before the corresponding data were collected, and that the file's
timestamp and SHA-256 hash are recorded in the project repository, so that
the order of prediction and data is documented; these are local records, not
a third-party time stamp. No external registry (such as OSF) was used. Three items were
pre-registered: (i)~the $\astar = 0.435 \pm 0.03$ prediction and the decision
criteria of the main collective experiment, before any collective episode
(2026-08-23); (ii)~the three-hypothesis prediction table of the polarity
discrimination experiment, before its data collection (2026-08-25);
(iii)~the claim-selection rule of the eight-claim calibration campaign,
written into the campaign instructions before its data were collected
(2026-08-27). The single-agent measurements preceded the collective
experiments. The claim-resolved predictions of
Sections~4.2--4.4 of the main text were computed after the
collective experiments, with the estimation procedure described under
Statistics, using only the single-agent data for parameter fitting; they were
not prospectively registered. The one adaptive element (second-pass allocation of episodes)
followed rules fixed in the pre-registered design.

\end{document}